\documentclass[11pt, a4paper]{article}

\usepackage{setspace}
\usepackage{microtype}
\usepackage{bm}
\usepackage{upgreek}
\usepackage[numbers]{natbib}
\usepackage{rotating}
\usepackage{graphicx}
\usepackage{verbatim}
\usepackage{wrapfig}
\usepackage[margin=2.5cm]{geometry}
\usepackage{amsmath}
\usepackage{placeins}
\usepackage{url}

\usepackage{fancyhdr}
\usepackage{color}
\usepackage{amsthm}
\usepackage{amsmath, amssymb}
\usepackage{mathtools}
\usepackage{mathrsfs}
\usepackage{mathabx}
\usepackage{longtable}
\usepackage{lscape}
\usepackage{booktabs}
\usepackage{multicol}
\usepackage{pdfpages}
\usepackage{siunitx}
\usepackage{graphicx}
\usepackage[shortlabels]{enumitem}
\usepackage{doi}

\usepackage{pgfplots}
\pgfplotsset{compat=1.18}

\usepackage[affil-it]{authblk}

\usepackage{caption}
\usepackage{subcaption}

\usepackage{bbm}
\usepackage{nameref}

\usepackage{tikz}
\usetikzlibrary{positioning}
\tikzset{>=stealth}
\usetikzlibrary{decorations.markings}
\usetikzlibrary{patterns}
\usetikzlibrary{math}

\usepackage{hyperref}
\hypersetup{
  hidelinks,
  pdftitle={First-order phase transition in spin-imbalanced BCS theory},
  pdfauthor={Marius Lemm and Barbara Roos}
}

\usepackage{chngcntr}
\counterwithout{table}{section}

\newtheorem{all}{Theorem}[section]

\theoremstyle{plain}
\newtheorem{lemma}[all]{Lemma}
\newtheorem{thm}[all]{Theorem}

\newtheorem{prop}[all]{Proposition}

\theoremstyle{definition}

\newtheorem{rem}[all]{Remark}
\newtheorem{asptn}[all]{Assumption}

\newcommand{\BC}{{\mathbb{C}}}

\newcommand{\BR}{{\mathbb{R}}}
\newcommand{\BS}{{\mathbb{S}}}

\newcommand{\CC}{{\mathcal{C}}}

\newcommand{\dd}{{\mathrm{d}}}

\newcommand{\tr}{\mathrm{Tr}\ }

\DeclareMathOperator\sech{sech}

\newcommand{\com}[1]{}

\newcommand{\R}{\mathbb R}
\newcommand{\C}{\mathbb C}
\newcommand{\Sph}{\mathbb S}
\newcommand{\F}{\mathcal F}
\newcommand{\Vop}{\mathcal V}
\newcommand{\Tr}{\operatorname{Tr}}
\newcommand{\spec}{\sigma}
\newcommand{\op}{\mathrm{op}}
\newcommand{\sgn}{\operatorname{sgn}}
\newcommand{\1}{\mathbf 1}
\newcommand{\mub}{\overline{\mu}}
\newcommand{\dm}{\delta_\mu}

\newcommand{\llin}{\lambda_{\mathrm{lin}}}
\newcommand{\lfo}{\lambda_{\mathrm{fo}}}
\newcommand{\Da}{L_1^2(\BR^d)}
\newcommand{\Wop}{|V|^{1/2}}

\numberwithin{equation}{section}
\begin{document}
\title{A First-Order Phase Transition in Spin-Imbalanced BCS Theory of Superconductivity}
\author[1]{Marius Lemm\thanks{marius.lemm@uni-tuebingen.de}}
\author[2]{Barbara Roos\thanks{barbara.roos@gssi.it}}
\affil[1]{Department of Mathematics, University of T\"ubingen, 72076 T\"ubingen, Germany}
\affil[2]{Gran Sasso Science Institute, Viale Francesco Crispi 7, 67100 L’Aquila, Italy}

\date{\today}

\maketitle

\begin{abstract}
We study a translation-invariant BCS functional of superconductivity for a system of spin-$1/2$ fermions with spin-dependent chemical potentials, $\mu_+\neq\mu_-$. Numerical work of Caldas and Mota and analytical work of Freiji, Hainzl, and Seiringer has suggested an unusual first-order phase transition in spin-imbalanced BCS models, in which the normal state becomes globally unstable while remaining locally stable. Here, we prove that a first-order phase transition occurs in such spin-imbalanced BCS functionals. For each sufficiently small fixed spin imbalance and sufficiently low temperature, there is a coexistence coupling $\lfo$ strictly below the linear-stability threshold $\llin$ of the normal state at which the Cooper pair density exhibits a nonzero jump and the minimized energy has a kink. The methodological insight is a coercivity estimate obtained by decomposing the BCS energy minimization into two parts, where the ``inner'' problem minimizes over the one-particle density matrices at fixed pairing and can be exactly solved by a Legendre--Fenchel transform.
\end{abstract}

\section{Introduction}

In the mathematical analysis of the BCS functional of superconductivity, a central role is played by the normal
state and its stability. The normal state minimizes the BCS functional among states with zero pairing density. If it is globally unstable, a minimizer with non-zero pairing exists and the physical system becomes superconducting.  The onset of superconductivity as the temperature is varied is commonly governed by a  second-order phase transition in which the pairing density of competing minimizers to the normal state vanishes continuously at the critical temperature. Hence, the linear stability criterion (positive definite Hessian operator) of the normal state plays a crucial role in describing the second-order phase transition. In particular, in the translation-invariant spin-balanced model, linear stability characterizes the critical temperature \cite[Theorem~1]{HainzlHamzaSeiringerSolovej2008}. The Hessian can become a subtle operator in itself, especially in the presence of external fields \cite{FrankHainzlSeiringerSolovej2016} and it plays a central role in the celebrated rigorous derivation of Ginzburg--Landau (GL) theory near the critical temperature \cite{FrankHainzlSeiringerSolovej2012}. See \cite{HainzlSeiringer2016} for a survey of the mathematical properties of the BCS functional of superconductivity.
In domains with boundary, a rigorous derivation of GL theory remains open. An exception is at zero temperature, where \cite{FrankLemmSimon2017} derived an effective Gross-Pitaevskii functional for domains with boundary in the Bose--Einstein condensation regime.
For a derivation of GL theory near the superconducting transition temperature, a key issue is to establish its second-order nature in the presence of a boundary. 
This question is complicated by the possibility of boundary superconductivity 
\cite{HainzlRoosSeiringer2022,RoosSeiringer2025halfspaces, RoosSeiringer2025corner,SamoilenkaBabaev2020}. This  motivates the broader need to develop new tools to analyze the nature of different phase transitions  in BCS models. 

In a spin-imbalanced system, this usual picture of a second-order phase transition is expected to break down, e.g., in the context of Fulde--Ferrell--Larkin--Ovchinnikov (FFLO) superconductors \cite{AgterbergYang2001,BuzdinKachkachi1997,CasalbuoniNardulli2004,CaoEtAl2026,CombescotMora2002,MakiTsuneto1964,MatsudaShimahara2007,ParishMarchettiLamacraftSimons2007,ZdybelJakubczyk2020}. In particular, it has been predicted in the physics literature that the paired phase can become
globally favorable while the normal state still has a positive Hessian and that the pairing density of the newly arising competitor is non-zero at the coexistence point, leading to a kink in the energy, i.e., a first-order phase transition \cite{BuzdinKachkachi1997,caldas_temperature_2008,Sarma1963}. In particular, line III in Figure~2 of \cite{caldas_temperature_2008} records a numerical prediction for the occurrence of this phase transition in a spin-imbalanced BCS functional with delta interaction. In this paper, we rigorously confirm that such a first-order phase transition occurs for a broad class of interaction potentials in the regime of small spin imbalance and low temperature. The precise statement of our main result can be found in Theorem~\ref{thm:main} below. In the mathematical physics literature, the spin-imbalanced BCS functional was studied by Freiji, Hainzl, and Seiringer \cite{FreijiHainzlSeiringer2012}.  
The techniques of \cite{FreijiHainzlSeiringer2012}, in particular, gave a spectral characterization of the existence theory of solutions to the nonlinear gap equation (i.e., of critical points of the BCS free energy functional) with non-zero pairing. The absence of critical points can then be used to exclude the possibility of superconductivity in suitable parameter regimes. However, these techniques are less suited to identify situations in which the normal state is locally stable but globally unstable --- or when the normal state is not the unique minimizer, which is the case at a first-order phase transition where a normal and superconducting minimizer coexist. 

The present paper uses different optimization ideas, related to Legendre-Fenchel transforms, to confirm the existence of a first-order phase transition in the coupling constant, as we describe below. The link to the temperature-dependent picture of \cite{FreijiHainzlSeiringer2012} is explained in Appendix \ref{app:temperature}.

\subsection*{Main result and its significance}

Our main theorem separates two coupling thresholds that coincide in the usual
second-order BCS picture.  Fixing a sufficiently small imbalance $\dm>0$ and
then taking the temperature sufficiently low, we prove that there is a
coexistence coupling $\lfo(T,\dm)$ strictly below the
linear-stability threshold $\llin(T,\dm)$ of the normal
state:
\[
 0<\lfo(T,\dm)
 <\llin(T,\dm).
\]
More precisely, the stability transitions are as follows:
\begin{itemize}
\item For $0\leq\lambda<\lfo$, the normal state is the unique global minimizer.
\item At $\lambda=\lfo$, it coexists with a superconducting global minimizer.
\item For $\lfo<\lambda<\llin$, the superconducting phase has lower energy, but the normal state remains a strict local minimizer.
\item At $\lambda=\llin$, the normal state is no longer a local minimizer.
\item For $\lambda>\llin$, the Hessian has negative spectrum and the normal state is locally unstable.
\end{itemize}
Thus the transition is first order in a variationally
precise sense: the order parameter jumps at coexistence, metastability is
present on one side, and the minimized energy has a kink as a function of the
coupling.

The point of the theorem is that none of these global conclusions follows
from the Hessian alone.  The linearized problem locates $\llin$. We show that at this threshold, the fourth-order perturbative coefficient along the critical mode is negative, which leads to global instability of the normal state for couplings slightly below $\llin$.  To turn
this local information into a coexistence
theorem, we rewrite the full BCS minimization as an exact constrained search
over pairing densities, solve the inner problem by Legendre--Fenchel duality,
and use the resulting nonlinear coercivity together with compactness to show
that the coexistence threshold is attained by a nonzero state.
This also explains the relation with the phase diagram of Caldas and Mota:
the equality-of-minima locus is the analogue of their line III, whereas the
Hessian-zero locus corresponds to the lower-temperature branch of their line II; see Appendix~\ref{app:temperature}.

\begin{figure}
\centering 

\begin{tikzpicture}[scale=0.95]
\begin{axis}[
    width=10cm,
    height=10cm,
    axis lines=left,
    xlabel={$\Gamma^{[\alpha]}$},
    ylabel={$\mathfrak F(\Gamma)$},
    xmin=-0.2, xmax=1.3,
    ymin=-2.2, ymax=1.4,
    xtick={0,1.14},
    xticklabels={$\Gamma_0$,$\Gamma^{[\alpha_*]}$},
    ytick=\empty,
    samples=300,
    clip mode=individual,
    xlabel style={at={(axis description cs:1,0)}, anchor=west},
    ylabel style={at={(axis description cs:0,1)}, anchor=south},
]

\addplot[black, thick, domain=-0.2:1.3]
    {x^6 - 2.5*x^4 + 2*x^2 + 0.6};

\addplot[black, thick, domain=-0.2:1.3]
    {x^6 - 2.6*x^4 + 1.69*x^2 + 0.35};

\addplot[black, thick, domain=-0.2:1.3]
    {x^6 - 2.6*x^4 + 1.25*x^2 + 0.1};

\addplot[black, thick, domain=-0.2:1.3]
    {x^6 - 2*x^4 - 0.1};

\addplot[black, thick, domain=-0.2:1.3]
    {x^6 - 1.7*x^4 - 0.7*x^2 - 0.4};

\node[anchor=west, text=black]
    at (axis cs:1.34,1.30) {(i) $\lambda<\lfo$};

\node[anchor=west, text=black]
    at (axis cs:1.34,0.61) {(ii) $\lambda=\lfo$};

\node[anchor=west, text=black]
    at (axis cs:1.34,-0.39) {(iii) $\lfo<\lambda<\llin$};

\node[anchor=west, text=black]
    at (axis cs:1.34,-0.99) {(iv) $\lambda=\llin$};

\node[anchor=west, text=black]
    at (axis cs:1.34,-1.61) {(v) $\lambda>\llin$};

\end{axis}
\end{tikzpicture}

\caption{Qualitative sketch of the energy landscape at different couplings, with vertical offsets for clarity. 
Regime (i) $\lambda<\lfo$: normal state unique minimizer, (ii) $\lambda=\lfo$: normal and paired state  both global minimizers, (iii) $\lfo<\lambda  <\llin$: paired global minimizer while normal state strict local minimizer, (iv) $\lambda=\llin$: normal state locally unstable while Hessian nonnegative, (v) $\lambda>\llin$: normal state locally unstable due to negative Hessian.
}\label{fig:fig1}
\end{figure}

\section{Setup and main results}
\label{sec:model-main}
\subsection{The model}
As in \cite{FreijiHainzlSeiringer2012}, we consider a system of spin-$1/2$ fermions in dimension $d\in\{1,2,3\}$ at temperature $T> 0$.
We assume that for spin-up and spin-down states there are two different chemical potentials $\mu_+,\mu_-\in\R$, respectively.
We denote the spin imbalance and the average chemical potential by
\[
\delta_\mu=\frac{\mu_+-\mu_-}{2}, \quad \mub=\frac{\mu_++\mu_-}{2}.
\]
We assume throughout that $\bar\mu>0$.
This results in the two spin states being unequally populated.
We assume that the effective interaction between opposite spin states is $\lambda V$, where $V:\R^d\to\R$ is real and even and $\lambda$ is a positive coupling constant.

As explained in Appendix~\ref{sec:model_justification}, we consider measurable matrix-valued states $\Gamma:\R^d\to\C^{4\times4}$ of the form 
\begin{equation}\label{eq:block-state}
\Gamma(p)=\begin{pmatrix} \gamma_+(p)&0&0& \alpha(p)\\ 
0 & \gamma_{-}(p) & - \alpha(p)& 0\\
0 & -\overline{ \alpha(p)} & 1-\gamma_+(p)& 0\\
\overline{ \alpha(p)}& 0 &0 & 1-\gamma_{-}(p) \end{pmatrix} \qquad \text{such that}
 \qquad 0\leq\Gamma(p)\leq1
\end{equation}
for almost every $p\in\R^d$, where $\gamma_\pm:\R^d\to[0,1]$ and $\alpha:\R^d\to\C$ are measurable.
The BCS functional of superconductivity is
\begin{equation}\label{eq:BCSfunctional}
\boxed{\mathfrak{F}(\Gamma)= \int_{\BR^d}\Big( \frac{1}{2}(p^2-\mu_+)\gamma_+(p)+\frac{1}{2} (p^2-\mu_-)\gamma_-(p)\Big) \dd p - \frac{T}{2} S(\Gamma)-\lambda\langle \alpha,\Vop \alpha \rangle}
\end{equation}
Here $S$ denotes the entropy
\begin{equation}
    S(\Gamma)=-\frac{1}{2}\int_{\BR^d} \tr_{\BC^4}[ \Gamma(p)\ln \Gamma(p) +(1- \Gamma(p))\ln (1-\Gamma(p) )]\dd p
\end{equation}
and $\Vop:=\F V\F^{-1}$ is the interaction in momentum space, understood as a quadratic form on $\Da:=L^2(\R^d,(1+p^2)\dd p;\C)$.
The Fourier transform $\F:L^2(\R^d)\to L^2(\R^d)$ is given by
\[
 (\F f)(p)=(2\pi)^{-d/2}\int_{\R^d}e^{-ip\cdot x}f(x)\dd x,\qquad f\in L^1(\R^d)\cap L^2(\R^d),\]
extended to $L^2(\R^d)$ by continuity.
 We assume
\begin{equation}\label{eq:V-assumptions}
 V\in L^1(\R^d),
 \qquad
 \begin{cases}
 \text{no additional assumption},&d=1,\\
 V\in L^{1+\varepsilon}(\R^2)\text{ for some }\varepsilon>0,&d=2,\\
 V\in L^{3/2}(\R^3),&d=3.
 \end{cases}
\end{equation}
These assumptions are standard in the analysis of the BCS model, compare
\cite{HainzlHamzaSeiringerSolovej2008,HenheikLauritsenRoos2024}, and imply that the BCS functional is well-defined and has a minimizer on the domain
\begin{equation}\label{eq:domain}
  \mathcal{D}:=\left\{\Gamma:\R^d\to\C^{4\times4}\ \text{measurable, of the form }\eqref{eq:block-state}\;:\;\gamma_\pm\in L^1(\R^d,(1+p^2)\dd p)\right\}.
\end{equation}
Note that the condition $0\leq \Gamma\leq 1$ implies $0\leq \gamma_\pm(p) \leq 1$ and $|\alpha(p)|^2\leq \min\{\gamma_-(p)(1-\gamma_+(p)), \gamma_+(p)(1-\gamma_-(p))\}$ for almost every $p$.
In particular, $|\alpha|\leq1/2$ almost everywhere and
$\alpha\in\Da$.

\subsection{The normal state and its Hessian}

The normal state is
\begin{equation}\label{eq:normal-state}
 \Gamma_0(p)=\operatorname{diag}
 \left(n_+(p),n_-(p),1-n_+(p),1-n_-(p)\right),
 \qquad
 n_\pm(p)=\frac1{1+e^{(p^2-\mu_\pm)/T}}.
\end{equation}
The normal state is the unique minimizer of $\mathfrak{F}$ at $\lambda=0$, and hence the unique minimizer among states with $\alpha=0$ for every $\lambda\geq0$. Local minimality  is understood in the topology induced by the metric
\[
 d( \Gamma,\Gamma')=\lVert \Gamma-\Gamma'\rVert_{\rm HS}=\left(\int_{\R^d}\Tr_{\C^4}|\Gamma(p)-\Gamma'(p)|^2\dd p\right)^{1/2}.
\]
The linear-stability threshold $\llin$ is determined through the local stability of the normal state.
The normal state is always a critical point of the BCS functional, thus the relevant quantity to consider is the Hessian.
As shown in \cite[Proof of Theorem 1]{FreijiHainzlSeiringer2012}, with the notation
\[
\xi:\R^d\to\R,\qquad\xi(p)=|p|^2-\mub
\]
the Hessian at the normal state with respect to variations in $\alpha$ is $K_{T,\delta_\mu}-\lambda \Vop$, where
\begin{equation}\label{eq:K-def}
 K_{T,\dm}(p)=
 \frac{2|\xi(p)|}
 {\tanh\!\left(\frac{|\xi(p)|+\dm}{2T}\right)
 +\tanh\!\left(\frac{|\xi(p)|-\dm}{2T}\right)},
\end{equation}
with $K_{T,\dm}(p)=2T\cosh^2(\dm/(2T))$ at $\xi(p)=0$, by continuity. We use the same notation for the corresponding multiplication operator on $L^2(\R^d)$.  
To see that the operator $K_{T,\delta_\mu}-\lambda \Vop$ is well-defined, note that for fixed $T>0$ and $\dm>0$, there are constants $0<c<C<\infty$ such that
\begin{equation}\label{eq:K-growth}
 c(1+p^2)\leq K_{T,\dm}(p)\leq C(1+p^2).
\end{equation}
For $\alpha\in\Da$, the interaction form is
\[
 \langle\alpha,\Vop\alpha\rangle
 :=\int_{\R^d}V(x)|\F^{-1}\alpha(x)|^2\dd x.
\]
It has the factorization 
\begin{equation}\label{eq:interaction-form-definition}
 \langle \alpha,\Vop \alpha\rangle
 =\left\langle K_{T,\dm}^{1/2}\alpha,
 L_{T,\dm}K_{T,\dm}^{1/2}\alpha\right\rangle,
 \qquad
 L_{T,\dm}:=K_{T,\dm}^{-1/2}\Vop K_{T,\dm}^{-1/2}.
\end{equation}
It follows from Lemma~\ref{lem:compact-sandwich} below that $L_{T,\dm}$ is compact and
self-adjoint.  
Hence, the interaction is infinitesimally form bounded with respect to $K_{T,\dm}$, see for instance \cite[Lemma 6.26]{teschl_mathematical_2014}, and $K_{T,\dm}-\lambda \Vop$ is well-defined by the KLMN theorem \cite[Theorem 6.24]{teschl_mathematical_2014}.
For parameters for which
$b_{T,\dm}:=\max\spec L_{T,\dm}>0$, the linear-stability threshold is 
\begin{equation}\label{eq:lambda-lin}
 \llin(T,\dm):=b_{T,\dm}^{-1}.
\end{equation}
By the Birman--Schwinger principle, $\llin$ is precisely the coupling such that $\inf\sigma(K_{T,\dm}-\llin \Vop)=0$ \cite[Section III.C]{HainzlSeiringer2016}.
For $\lambda>\llin$ there is negative spectrum and thus the normal state is unstable.

\subsection{The threshold coupling constant for the first-order transition}
To define the second threshold, we introduce the free-energy difference 
\begin{equation}\label{eq:functional}
 \mathcal E_{T,\dm,\lambda}(\Gamma)
 :=\mathfrak{F}(\Gamma)-\mathfrak{F}(\Gamma_0)=\int_{\R^d}\left(\ell_{T,p}(\Gamma(p))
       -\ell_{T,p}(\Gamma_0(p))\right)\dd p
 -\lambda\langle \alpha,\Vop \alpha\rangle,  
\end{equation}
where $\ell_{T,p}$ is the pointwise free-energy density
\begin{equation}\label{eq:ell-def}
 \ell_{T,p}(\Gamma)
 =\frac12(p^2-\mu_+)\gamma_+
  +\frac12(p^2-\mu_-)\gamma_-
  +\frac T4\Tr_{\C^4}\!\left[
    \Gamma\ln\Gamma+(1-\Gamma)\ln(1-\Gamma)\right].
\end{equation}
Define $G_{T,\dm}:\R^d\times\C\to \mathbb R\cup \{+\infty\}$ by
\begin{equation}\label{eq:G-constrained-definition}
 G_{T,\dm}(p,z)
 :=\inf_{\substack{0\leq\Gamma\leq1,\ \Gamma\ \mathrm{of\ the\ form}
                         \ \eqref{eq:block-state}\\
                    \Gamma_{14}=z}}
 \bigl(\ell_{T,p}(\Gamma)-\ell_{T,p}(\Gamma_0(p))\bigr).
\end{equation}
Here the infimum is over matrices in $\C^{4\times4}$. The matrix constraint implies $|z|\leq1/2$; we set
$G_{T,\dm}(p,z)=+\infty$ when $|z|>1/2$. 
With this, we define the crucial quantity in this work, the \textit{threshold coupling constant for the first-order transition}
\begin{equation}\label{eq:lambda-fo}
 \boxed{\lfo(T,\dm)
 :=\inf_{\substack{\alpha\in \Da,\ |\alpha|\leq1/2\\
                    \langle \alpha,\Vop \alpha\rangle>0}}
 \frac{\displaystyle\int_{\R^d}G_{T,\dm}(p,\alpha(p))\dd p}
      {\langle \alpha,\Vop \alpha\rangle}}
\end{equation}
Using that the normal state is the minimizer of $\mathfrak{F} $ among states with $\langle \alpha,\Vop \alpha\rangle\leq 0$ and the definition of $G_{T,\dm}$ it is immediate that
\begin{multline}\label{eq:two-step-minimization-inequality}
 \inf_{\Gamma\in \mathcal{D}}  \mathcal E_{T,\dm,\lambda}(\Gamma)
 =\inf_{\substack{\alpha \in \Da,\ |\alpha|\leq1/2 \\
                    \langle \alpha,\Vop \alpha\rangle\geq 0}}
 \inf_{\substack{\Gamma\in \mathcal{D}\\
                    \Gamma_{14}=\alpha }}
                    \mathcal E_{T,\dm,\lambda}(\Gamma)\\
 \geq \inf_{\substack{\alpha \in \Da,\ |\alpha|\leq1/2\\ \langle \alpha,\Vop \alpha\rangle\geq 0}}
\left[\int_{\R^d}G_{T,\dm}(p,\alpha(p))\dd p
-\lambda \langle \alpha,\Vop \alpha\rangle \right].
\end{multline}
For $\lambda \leq \lfo$, the bracketed expression is nonnegative and thus the normal state is a global minimizer of the BCS functional. 
In Lemma~\ref{lem:fixed-gap-duality} we shall show that the above is in fact an equality.
It then follows that for $\lambda > \lfo$, when the bracketed expression attains negative values, the normal state is not the global minimizer.

\subsection{Assumption on the interaction potential}
We shall consider a regime where both spin imbalance $\delta_\mu$ and temperature $T$ are small. 
As in the usual, spin-balanced case \cite{FHNS2007}, the restriction of the interaction to the Fermi surface will be important in this regime.
To define this operator rigorously, set
\[
 (Jf)(x):=\sgn(V(x))f(x),\qquad f\in L^2(\R^d).
\]
For $s>0$, define $\mathcal T_s:L^2(\R^d)\to L^2(\Sph^{d-1})$ by
\begin{equation}\label{eq:T-s-def}
 (\mathcal T_s f)(\omega)
 =(2\pi)^{-d/2}\int_{\R^d}e^{-i\sqrt{s}\,\omega\cdot x}\Wop (x)f(x)\dd x,
 \qquad \omega\in\Sph^{d-1}.
\end{equation}
Here $\Sph^{d-1}=\{\omega\in\R^d:|\omega|=1\}$ carries surface measure (counting measure when $d=1$). The operator is Hilbert--Schmidt from $L^2(\R^d)$ to $L^2(\Sph^{d-1})$, since
\begin{equation}\label{eq:T-s-HS}
 \|\mathcal T_s\|_{\mathrm{HS}}^2
 =(2\pi)^{-d}|\Sph^{d-1}|\|V\|_1.
\end{equation}
The Fermi-surface interaction is the compact self-adjoint operator
\begin{equation}\label{eq:surface-op}
 \Vop_{\mub}
 :=\mub^{(d-2)/2}\mathcal T_{\mub}J\mathcal T_{\mub}^*
 \quad\text{on }L^2(\Sph^{d-1}).
\end{equation}
We can now formulate our assumption on $V$.

\begin{asptn}\label{as:V}
   Let $\bar\mu>0$. Suppose that $V$ satisfies \eqref{eq:V-assumptions} and suppose that the largest eigenvalue of \(\Vop_{\mub}\) is strictly positive and simple.  
\end{asptn}

\subsection{Main result}
Our main result is the following.
\begin{thm}[First-order transition in the coupling]\label{thm:main} Let $\bar\mu>0$.
Suppose that $V$ satisfies Assumption \ref{as:V}.
Then, there exists $\delta_{\mu,0}\in(0,\mub)$ such that,
for every $0<\dm<\delta_{\mu,0}$, there is $T_0(\dm)>0$ with the following
property.  For every $0<T<T_0(\dm)$,
\begin{equation}\label{eq:threshold-ordering-main}
 0<\lfo(T,\dm)
   <\llin(T,\dm)<\infty.
\end{equation}
The BCS functional has a global minimizer for every $\lambda\geq0$.  Moreover:
\begin{enumerate}[label=\textup{(\roman*)}]
\item If $0\leq\lambda<\lfo(T,\dm)$, then $\Gamma_0$ is
the unique global minimizer.
\item At $\lambda=\lfo(T,\dm)$, the normal state and a
nonnormal state $\Gamma^{[\alpha_*]}$ coexist as distinct global minimizers, with
\begin{equation}\label{eq:coexistence-main}
 \alpha_*\in\Da\setminus\{0\},\qquad \langle \alpha_*,\Vop \alpha_*\rangle>0,\qquad
 \int_{\R^d}G_{T,\dm}(p,\alpha_*(p))\dd p
 =\lfo(T,\dm)\langle \alpha_*,\Vop \alpha_*\rangle.
\end{equation}
\item If
$\lfo(T,\dm)<\lambda<\llin(T,\dm)$,
then every global minimizer is paired, while $\Gamma_0$ remains a strict local minimizer, that is, for each
$0\leq\bar\lambda<\llin(T,\dm)$ there are
$\rho,c_{\mathrm{loc}}>0$ such that, uniformly for
$0\leq\lambda\leq\bar\lambda$,
\begin{equation}\label{eq:L2-local-coercivity-main}
 \mathcal E_{T,\dm,\lambda}(\Gamma)
 \geq c_{\mathrm{loc}}\langle \alpha,K_{T,\dm}\alpha\rangle
\end{equation}
for every $\Gamma\in\mathcal D$ with $\|\alpha\|_2<\rho$. 
In particular, every nonnormal minimizer with $0\leq\lambda\leq\bar\lambda$ satisfies $\|\alpha\|_2\geq\rho$. Taking $\bar\lambda>\lfo$ gives this bound also for the coexistence state $\alpha_*$. 
Hence, the pairing density of the minimizer jumps at $\lfo$.
\item At $\lambda=\llin(T,\dm)$ the normal state is not a
local minimizer. 
\item For $\lambda>\llin(T,\dm)$ the normal state is locally unstable since the Hessian has negative spectrum.
\end{enumerate}
Finally,
with
\begin{equation}\label{eq:minimized-energy-main}
 e_{T,\dm}(\lambda):=\inf_{\Gamma\in\mathcal D}
 \mathcal E_{T,\dm,\lambda}(\Gamma),
\end{equation}
one has $e_{T,\dm}(\lambda)=0$ for
$0\leq\lambda\leq\lfo(T,\dm)$ and, for $h>0$,
\begin{equation}\label{eq:kink-main}
 e_{T,\dm}(\lfo+h)
 \leq-h\langle \alpha_*,\Vop \alpha_*\rangle<0.
\end{equation}
The function $e_{T,\dm}$ is concave, and its one-sided derivatives at
coexistence satisfy
\begin{equation}\label{eq:one-sided-slopes-main}
 e'_{T,\dm,-}(\lfo)=0,
 \qquad
 e'_{T,\dm,+}(\lfo)
 \leq-\langle \alpha_*,\Vop \alpha_*\rangle<0.
\end{equation}
\end{thm}

Theorem \ref{thm:main} is proved in Section \ref{eq:conclusion}. The two statements in \eqref{eq:one-sided-slopes-main} imply that the minimized energy has a first-order kink, which establishes the first-order phase transition. 

In physical terms, the assumption on $V$ that the largest eigenvalue of $\mathcal V_{\bar\mu}$ is strictly positive and simple selects a unique leading pairing channel.
If $V$ is additionally assumed to be radial and $d\in\{2,3\}$, this leading channel is the s-wave channel.
In principle, one can also study degenerate situations, such as d-wave superconductivity, and it would be interesting to explore the results of \cite{FrankLemm2016} in the spin-imbalanced setting.

\subsection{Proof strategy}\label{sec:overview}
The proof rests on three propositions. Proposition~\ref{prop:variational-first-order} gives coexistence and the jump once a negative-energy trial state exists at $\llin$. Proposition~\ref{prop:BS} controls the critical eigenvector, and Proposition~\ref{prop:negative-quartic} uses it to construct such a trial state. Together they imply Theorem~\ref{thm:main}. We now explain the variational mechanism and the estimates behind these propositions. 
\paragraph*{Step 1: $\lfo$ determines global stability of the normal state}
It follows from \eqref{eq:two-step-minimization-inequality} that the normal state is a global minimizer if $\lambda\leq \lfo$.
Provided that \eqref{eq:two-step-minimization-inequality} is actually an equality, the minimizer is paired for $\lambda > \lfo$.

For $\alpha\in\Da$, let
\begin{equation}\label{eq:R-definition}
 \mathcal R_{T,\dm,\lambda}(\alpha)
 :=\int_{\R^d}G_{T,\dm}(p,\alpha(p))\dd p
   -\lambda\langle \alpha,\Vop \alpha\rangle.
\end{equation}
The definition of $G_{T,\dm}$ makes this an extended-real functional,
with $\mathcal R_{T,\dm,\lambda}(\alpha)=+\infty$ if $|\alpha|>1/2$ on a set of positive measure.
We write $\mathcal R_\lambda(\alpha)$ when $T$ and $\dm$ are fixed.

For $\alpha\in \Da$ with $\lVert \alpha\rVert_\infty\leq 1/2$, Lemma~\ref{lem:fixed-gap-duality} shows that there is a measurable admissible state $\Gamma^{[\alpha]}$ satisfying
\[
 \ell_{T,p}(\Gamma^{[\alpha]}(p))-\ell_{T,p}(\Gamma_0(p))
 =G_{T,\dm}(p,\alpha(p))
 \quad\text{for almost every }p.
\]
Consequently, 
$\mathcal E_{T,\dm,\lambda}(\Gamma^{[\alpha]})
=\mathcal R_{T,\dm,\lambda}(\alpha)$, that is, \eqref{eq:two-step-minimization-inequality} is in fact an equality.
In particular, $\lfo$ determines the global stability of the normal state.

\paragraph*{Step 2:  $\llin$ determines local stability of the normal state}
Since $K_{T,\dm}>0$, for $\lambda >\llin$ we have $\inf \sigma(K_{T,\dm}-\lambda \Vop)<0$.
Hence, the normal state is locally unstable.
On the other hand, we shall prove in Proposition~\ref{prop:L2-local-stability} that for every $0\leq\overline{\lambda}<\llin$ there are $c,\rho>0$ such that for all $0\leq\lambda\leq\overline{\lambda}$ and $\alpha\in\Da$ with $|\alpha|\leq1/2$ almost everywhere and $\lVert\alpha\rVert_2<\rho$
\begin{equation}\label{eq:overview-coercivity}
\mathcal R_{T,\dm,\lambda}(\alpha)\geq c\langle \alpha,K_{T,\dm} \alpha\rangle.
\end{equation}
In particular, the normal state is stable against all perturbations with $\alpha$ small in $L^2$ norm.
(This is stronger than stability against perturbations of the form $t\alpha$ for small $t$ and fixed $\alpha$.)
Hence for $\lambda<\llin$ the normal state indeed is a local minimizer of the BCS functional in the topology specified above.

\paragraph*{Step 3: Global instability of the normal state at $\llin$ implies the main theorem}
Suppose that at the coupling $\llin$ there is a state with energy lower than the normal state. Then by Step 1, $\lfo<\llin$.
Proposition~\ref{prop:variational-first-order} explains how to conclude the main theorem under this assumption.
Compactness of the Birman--Schwinger operator and the pointwise constraint
$|\alpha|\leq1/2$ imply attainment at $\lfo$ by a nonzero
pairing density $\alpha_*\in\Da$.  Consequently
\[
 \int_{\R^d}G_{T,\dm}(p,\alpha_*(p))\dd p
 =\lfo(T,\dm)\langle\alpha_*,\Vop\alpha_*\rangle,
\]
so the normal and paired states are both global minimizers.
Coercivity \eqref{eq:overview-coercivity} separates the paired
minimizer from zero in $L^2$ and yields the order-parameter jump.

\paragraph*{Step 4: Proving local instability of the normal state at $\llin$}
 Let $\alpha_{T,\dm}\in\Da$ be the eigenfunction of $K_{T,\dm}-\llin \Vop$ corresponding to the zero eigenvalue.
With the normalization in \eqref{eq:alpha-critical}, we shall prove an expansion of the form
\begin{align}
 \mathcal R_{T,\dm,\lambda}(t \alpha_{T,\dm})
 &=   \langle \alpha_{T,\dm},(K_{T,\dm}-\lambda\Vop )\alpha_{T,\dm}\rangle |t|^2 \notag\\
 &\quad
 -2\llin(T,\dm)^4
   I_{T,\dm}(|\Vop \alpha_{T,\dm} |^4)|t|^4+O(|t|^6),
 \label{eq:blackboard-landau}
\end{align}
where $I_{T,\dm}$ is defined in \eqref{eq:I-def} and $I_{T,\dm}(|\Vop\alpha_{T,\dm}|^4)>0$ in the parameter regime of the theorem.
At $\lambda=\llin$ the quadratic term vanishes and the fourth-order term is negative. 
In particular, the normal state is not a minimizer at $\llin$.

The regime of low $T$ and $\dm$ allows us to determine the sign of $I_{T,\dm}(|\Vop\alpha_{T,\dm}|^4)$ analytically.
In particular, this regime corresponds to the well-studied weak-coupling regime \cite{FHNS2007,HainzlSeiringer2008,HenheikLauritsenRoos2024}, where we have good control of $\alpha_{T,\dm}$.

\begin{rem}
The sign of $I_{T,\dm}(|\Vop\alpha_{T,\dm}|^4)$ depends on the order in which $T$ and $\dm$ tend to zero.
Taking $\dm\to0$ first recovers the usual spin-balanced functional, where the quartic term is positive. This makes the derivation of the effective Ginzburg--Landau functional possible in that case \cite{FrankHainzlSeiringerSolovej2012}.
We shall take $T\to 0$ first, which leads to the negative quartic contribution.
Appendix~\ref{sec:simultaneous_scaling} discusses this in more detail for scalings of the form $\dm=\kappa T$.
\end{rem}

\subsection{Organization of the paper}

\label{sec:proof-outline}

Section~\ref{sec:model-main} defines the model, gives the main theorem and the main ideas of the proof.  
Section~\ref{sec:duality} derives the
Legendre representation, proves nonlinear coercivity and proves the main result assuming global instability of the normal state at $\llin$.
Section~\ref{sec:BS} provides the necessary analysis of the Hessian operator at low temperature and small spin imbalance.  
Section~\ref{sec:quartic} analyzes the sign of
the fourth-order coefficient and concludes the proof of the main result.
Appendix~\ref{sec:model_justification} records the quasi-free origin of the reduced model.
A comparison with the results in \cite{caldas_temperature_2008} is provided in Appendix~\ref{app:temperature}.
Appendix~\ref{sec:simultaneous_scaling} discusses different scalings of $T$ and $\dm$.

\section{The dual reduction and nonlinear coercivity}
\label{sec:duality}
At fixed pairing density $\alpha$, we minimize the free-energy difference $\mathcal{E}_{T,\dm,\lambda}$ over the admissible diagonal densities $\gamma_\pm$.
The interaction term is constant in this minimization.
We first solve the pointwise problem defining $G_{T,\dm}(p,z)$ in \eqref{eq:G-constrained-definition}.
The resulting bounds imply local coercivity below $\llin$ in Proposition~\ref{prop:L2-local-stability}.
Proposition~\ref{prop:variational-first-order} then uses this coercivity and compactness to establish coexistence, provided a negative-energy trial state exists at $\llin$.

For $p\in\R^d$ and $s\geq0$, set
\begin{equation}\label{eq:h-response}
 h_{T,\dm}(p,s)
 :=\frac{
 \tanh\!\left(\frac{\sqrt{\xi(p)^2+s}+\dm}{2T}\right)
 +\tanh\!\left(\frac{\sqrt{\xi(p)^2+s}-\dm}{2T}\right)}
 {2\sqrt{\xi(p)^2+s}},
\end{equation}
where the value at $(\xi,s)=(0,0)$ is defined by continuity as
$\sech^2(\dm/(2T))/(2T)$.  Then
$h_{T,\dm}(p,0)=K_{T,\dm}(p)^{-1}$.
For $\Delta\in\C$, introduce the Hermitian matrix
\begin{equation}\label{eq:H-Delta}
 H_\Delta(p)=
 \begin{pmatrix}
 p^2-\mu_+&0&0&-\Delta\\
 0&p^2-\mu_-&\Delta&0\\
 0&\overline\Delta&-p^2+\mu_+&0\\
 -\overline\Delta&0&0&-p^2+\mu_-
 \end{pmatrix}\in\C^{4\times4}
\end{equation}
and define the corresponding Gibbs state
\begin{equation}
    \Gamma_\Delta(p)=\left(1+e^{H_\Delta(p)/T}\right)^{-1}\in\C^{4\times4}.
\end{equation}

\begin{lemma}[Fixed-gap minimization]
\label{lem:fixed-gap-duality}
Let $T>0$ and $\dm\in \BR$.
For every $p\in\R^d$ and $z\in\C$ with $|z|\leq1/2$,
\begin{equation}\label{eq:G-def}
 G_{T,\dm}(p,z)
 =\sup_{D\geq0}\left\{
 D|z|-\frac14\int_0^{D^2}h_{T,\dm}(p,s)\dd s\right\}.
\end{equation}
If $|z|<1/2$, the supremum is attained at a unique
$D=|\Delta_p(z)|$, where $\Delta_p(z)\in\C$ is characterized by
\begin{equation}\label{eq:a-Delta}
 z=\frac12\Delta_p(z)
 h_{T,\dm}\bigl(p,|\Delta_p(z)|^2\bigr).
\end{equation}
Moreover, for $|z|\leq 1/2$, $z\mapsto G_{T,\dm}(p,z)$ is nonnegative, convex, continuous and we have
\begin{equation}\label{eq:G-integral}
 G_{T,\dm}(p,z)=\int_0^{|z|}\Delta_p(s)\dd s.
 \end{equation}
The map $(p,r)\mapsto\Delta_p(r)\in[0,\infty)$ is continuously differentiable on $\R^d\times[0,1/2)$ and
 \begin{equation}\label{eq:G-derivative}
 \partial_r^2G_{T,\dm}(p,0)=2K_{T,\dm}(p).
\end{equation}

If $\alpha \in \Da$ and $|\alpha|\leq 1/2$ almost everywhere, define
\begin{equation}\label{eq:optimizer-state}
 \Gamma^{[\alpha]}(p)
 :=\Gamma_{\Delta_p(\alpha(p))}(p)
\end{equation}
when $|\alpha(p)|<1/2$.  When $|\alpha(p)|=1/2$, set
\begin{equation}\label{eq:endpoint-state}
 \Gamma^{[\alpha]}(p):=
 \begin{pmatrix}
  1/2&0&0&\alpha(p)\\
  0&1/2&-\alpha(p)&0\\
  0&-\overline{\alpha(p)}&1/2&0\\
  \overline{\alpha(p)}&0&0&1/2
 \end{pmatrix}.
\end{equation}
The state $\Gamma^{[\alpha]} $ is measurable, admissible, has pairing density $\alpha$, and satisfies
\[
 \ell_{T,p}(\Gamma^{[\alpha]}(p))-\ell_{T,p}(\Gamma_0(p))
 =G_{T,\dm}(p,\alpha(p))
 \quad\text{for almost every }p.
\]
\end{lemma}
\begin{rem}
The dual representation \eqref{eq:G-def} is a pointwise Legendre--Fenchel
transform; see \cite{EkelandTemam1999} for the general framework and
\cite{LiebDensity1983} for a related constrained-optimization formulation in density
functional theory.  Here it separates the convex local term from the quadratic
interaction.
\end{rem}

\begin{rem}
For all $p$ with $|\alpha(p)|<1/2$, setting $z=\alpha(p)$ in \eqref{eq:a-Delta} expresses $\alpha(p)$ in terms of $\Delta(p)=\Delta_p(\alpha(p))$ as
\[
\alpha=\frac{1}{2} \Delta h_{T,\delta_\mu} (p,|\Delta|^2)
\]
The Euler-Lagrange equation for $\mathcal R_{T,\dm,\lambda}(\alpha)$ defined in \eqref{eq:R-definition} reads $\Delta=2\lambda\Vop\alpha$. Together, these relations yield the nonlinear BCS gap equation \cite[Eq.~(2.5)]{FreijiHainzlSeiringer2012}, which is stated solely in terms of $\Delta$, i.e.,
\[
 \Delta=\lambda\Vop\bigl(\Delta h_{T,\dm}(\cdot,|\Delta|^2)\bigr).
\]

\end{rem}

\begin{proof}
We first minimize at fixed gap, then invert the gap--pairing relation, and finally verify admissibility of the pointwise optimizer.

\emph{Step 1: minimization at fixed gap.}
Fix $p\in\R^d$.  For $\Delta\in\C$, define
\[
 \Phi_{p,\Delta}(\Gamma)
 :=\ell_{T,p}(\Gamma)-\operatorname{Re}(\overline\Delta\,\Gamma_{14}).
\]
Let
\begin{equation}\label{eq:q-def}
 q_{T,\dm}(p,\Delta)
 :=\inf_{0\leq\Gamma\leq1}
       \Phi_{p,\Delta}(\Gamma)-\Phi_{p,0}(\Gamma_0(p)),
\end{equation}
where the infimum is over matrices of the form \eqref{eq:block-state}.
For every such matrix with $\Gamma_{14}=z$, adding and subtracting the
linear term gives
\[
 \ell_{T,p}(\Gamma)-\ell_{T,p}(\Gamma_0(p))
 =\Phi_{p,\Delta}(\Gamma)-\Phi_{p,0}(\Gamma_0(p))
  +\operatorname{Re}(\overline\Delta z).
\]
Consequently,
\begin{equation}\label{eq:fixed-gap-supporting-bound}
 G_{T,\dm}(p,z)
 \geq q_{T,\dm}(p,\Delta)
       +\operatorname{Re}(\overline\Delta z).
\end{equation}
Equality holds whenever the minimizer in \eqref{eq:q-def} has $14$-entry
equal to $z$.

We now compute $q_{T,\dm}$.  
With $c_p=\frac12(p^2-\mub)$ independent of $\Gamma$,
\begin{equation}\label{eq:Phi-Gibbs-form}
 \Phi_{p,\Delta}(\Gamma)
 =\frac14\Tr(H_\Delta(p)\Gamma)
 +\frac T4\Tr\!\left[
 \Gamma\ln\Gamma+(1-\Gamma)\ln(1-\Gamma)\right]+c_p.
\end{equation}
By the finite-dimensional fermionic Gibbs variational principle (see for instance \cite[Section~IV.D]
{HainzlSeiringer2016})
its unique minimizer is the Gibbs state
\begin{equation}\label{eq:fixed-gap-Gibbs-state}
 \Gamma_\Delta(p)=\left(1+e^{H_\Delta(p)/T}\right)^{-1}.
\end{equation}
Indeed, for Hermitian matrices $A,B\in\C^{4\times4}$ with
$0\leq A\leq1$ and $0<B<1$, define the matrix relative entropy by
\[
 \mathcal H(A,B):=\Tr\!\left[
 A(\ln A-\ln B)+(1-A)(\ln(1-A)-\ln(1-B))\right].
\]
Since
$H_\Delta=T\bigl(\ln(1-\Gamma_\Delta)-\ln\Gamma_\Delta\bigr)$,
direct substitution in \eqref{eq:Phi-Gibbs-form} gives
\begin{equation}\label{eq:relative-entropy-identity}
 \Phi_{p,\Delta}(\Gamma)-\Phi_{p,\Delta}(\Gamma_\Delta)
 =\frac T4\mathcal H(\Gamma,\Gamma_\Delta)\geq0,
\end{equation}
where nonnegativity and the equality condition follow from Klein's inequality,
as in \cite[Section~IV.D]{HainzlSeiringer2016}.  Equality holds only if
$\Gamma=\Gamma_\Delta$.  The block formulas below show that $\Gamma_\Delta$
has the required form \eqref{eq:block-state}.  Thus \eqref{eq:fixed-gap-Gibbs-state}
is the unique minimizer in \eqref{eq:q-def}, and
$\Gamma_{\Delta=0}(p)=\Gamma_0(p)$.

We compute the entries of $\Gamma_\Delta$ explicitly.  Let
$e_1,\ldots,e_4$ be the standard basis of $\C^4$.  The subspaces
$\operatorname{span}\{e_1,e_4\}$ and $\operatorname{span}\{e_2,e_3\}$ are
invariant under $H_\Delta$.  In the indicated ordered bases, the corresponding
blocks are
\begin{equation}\label{eq:H-Delta-blocks}
 B_+=-\dm I_2+A_+,
 \qquad
 B_-=\dm I_2+A_-,
 \qquad
 A_+=\begin{pmatrix}\xi&-\Delta\\-\overline\Delta&-\xi\end{pmatrix},
 \quad
 A_-=\begin{pmatrix}\xi&\Delta\\\overline\Delta&-\xi\end{pmatrix},
\end{equation}
where $\xi=\xi(p)$.  Since
\(
 A_+^2=A_-^2=E_\Delta^2I_2,
  E_\Delta:=\sqrt{\xi^2+|\Delta|^2},
\)
the eigenvalues of $A_\pm$ are $\pm E_\Delta$, with spectral projections
\[
 P_{\pm}^{(\sigma)}
 =\frac12\left(I_2\pm\frac{A_\sigma}{E_\Delta}\right),
 \qquad \sigma\in\{+,-\}.
\]
The spectral-projection formulas apply when $E_\Delta>0$; the resulting Gibbs-state formulas are interpreted by continuity when $E_\Delta=0$.  
For $E_\Delta>0$, $c\in\R$, and any continuous function $F:\R\to\C$, finite-dimensional functional calculus gives
\begin{equation}\label{eq:two-by-two-functional-calculus}
 F(cI_2+A_\sigma)
 =F(c+E_\Delta)P_+^{(\sigma)}
  +F(c-E_\Delta)P_-^{(\sigma)}.
\end{equation}
We apply this with
\[
 F(x)=\frac1{1+e^{x/T}}
 =\frac12\left(1-\tanh\!\left(\frac{x}{2T}\right)\right)
\]
and abbreviate
\[
 t_\pm:=\tanh\!\left(\frac{E_\Delta\pm\dm}{2T}\right).
\]
For the first block, whose eigenvalues are $-\dm\pm E_\Delta$, one has
\[
 \tanh\!\left(\frac{B_+}{2T}\right)
 =t_-P_+^{(+)}-t_+P_-^{(+)}
 =\frac{t_--t_+}{2}I_2
  +\frac{t_++t_-}{2E_\Delta}A_+.
\]
Consequently,
\begin{equation}\label{eq:Gamma-first-block}
 F(B_+)
 =\left(\frac12+\frac{t_+-t_-}{4}\right)I_2
  -\frac{t_++t_-}{4E_\Delta}A_+.
\end{equation}
Similarly, the eigenvalues of the second block are $\dm\pm E_\Delta$, and
\[
 \tanh\!\left(\frac{B_-}{2T}\right)
 =t_+P_+^{(-)}-t_-P_-^{(-)}
 =\frac{t_+-t_-}{2}I_2
  +\frac{t_++t_-}{2E_\Delta}A_-.
\]
Hence
\begin{equation}\label{eq:Gamma-second-block}
 F(B_-)
 =\left(\frac12-\frac{t_+-t_-}{4}\right)I_2
  -\frac{t_++t_-}{4E_\Delta}A_-.
\end{equation}
Reading off the off-diagonal entry of \eqref{eq:Gamma-first-block}, we obtain
\begin{equation}
 (\Gamma_\Delta(p))_{14}
 =\frac{\Delta}{4E_\Delta}(t_++t_-)
 =\frac12\Delta h_{T,\dm}(p,|\Delta|^2).
 \label{eq:pairing-fixed-gap}
\end{equation}
Reading off the two upper diagonal entries gives
\begin{align}
 \gamma_{+,\Delta}
 &=\frac12+\frac{t_+-t_-}{4}
   -\frac{\xi}{4E_\Delta}(t_++t_-),\label{eq:gamma-plus-explicit}\\
 \gamma_{-,\Delta}
 &=\frac12-\frac{t_+-t_-}{4}
   -\frac{\xi}{4E_\Delta}(t_++t_-).
 \label{eq:gamma-minus-explicit}
\end{align}
These formulas also show directly that the minimizer depends continuously on $(p,\Delta)$.

\emph{Step 2: the dual representation and its regularity.}
Equality in \eqref{eq:fixed-gap-supporting-bound} holds if
$(\Gamma_\Delta(p))_{14}=z$.  We now find such a $\Delta$ for every $|z|<1/2$.
For $D\geq0$, set $E_D=\sqrt{\xi(p)^2+D^2}$ and define
\[
 f_p(D)=\frac12D h_{T,\dm}(p,D^2)
 =\frac{D}{4E_D}\left[
 \tanh\!\left(\frac{E_D+\dm}{2T}\right)
 +\tanh\!\left(\frac{E_D-\dm}{2T}\right)
 \right].
\]
The function in brackets is zero at $E_D=0$, positive for $E_D>0$, and
strictly increasing in $E_D$.  Since $D/E_D$ is nondecreasing, $f_p$ is
strictly increasing on $[0,\infty)$, with $f_p'(D)>0$ for $D>0$.
Moreover,
\[
 f_p(0)=0,\qquad \lim_{D\to\infty}f_p(D)=\frac12,
 \qquad f_p'(0)=\frac12h_{T,\dm}(p,0)>0.
\]

It follows from \eqref{eq:pairing-fixed-gap}, by matching the phase of $z$,
that for every $|z|<1/2$ there is a unique $\Delta_p(z)\in\C$ with
$(\Gamma_{\Delta_p(z)}(p))_{14}=z$.  Hence
\begin{equation}\label{eq:fixed-gap-supporting-equality}
 G_{T,\dm}(p,z)
 = \sup_{\Delta \in \BC} [ q_{T,\dm}(p,\Delta)
       +\operatorname{Re}(\overline\Delta z)].
\end{equation}
In particular, $\Delta_p(z)$ is the unique solution of \eqref{eq:a-Delta}.
To obtain an explicit expression for $q_{T,\dm}$ we compute 
\[
 \frac{\dd}{\dd t}
 \bigl(\Phi_{p,t\Delta}(\Gamma_{t\Delta}(p))
       -\Phi_{p,0}(\Gamma_0(p))\bigr)
 =-\operatorname{Re}\!\left(
 \overline\Delta(\Gamma_{t\Delta}(p))_{14}\right)
 =-\frac t2|\Delta|^2h_{T,\dm}(p,t^2|\Delta|^2),
\]
where we used that $\Gamma_{t\Delta}(p)$ is a minimizer of $\Phi_{p,t\Delta}$.
The expression in parentheses vanishes at $t=0$, so integration from $0$ to
$1$ and substituting $s=t^2 |\Delta|^2$ yields
\begin{equation}\label{eq:q-formula}
 q_{T,\dm}(p,\Delta)
 =-\frac14\int_0^{|\Delta|^2}h_{T,\dm}(p,s)\dd s.
\end{equation}
Combining \eqref{eq:q-formula} with
\eqref{eq:fixed-gap-supporting-equality} and optimizing the phase gives
\begin{equation}\label{eq:G-dual-lower-bound}
 G_{T,\dm}(p,z)
 =\sup_{D\geq0}\left\{
 D|z|-\frac14\int_0^{D^2}h_{T,\dm}(p,s)\dd s\right\}.
\end{equation}
In particular, for every $0\leq |z|<1/2$, the supremum is attained at $D=|\Delta_p(z)|$.

To prove \eqref{eq:G-integral}, it suffices by rotational symmetry to consider
real pairings $r\in[0,1/2)$.
The map $(p,r,D)\mapsto f_p(D)-r$ is
continuously differentiable and its derivative with respect to $D$ is strictly positive.  
By the implicit-function theorem, the map $(p,r)\mapsto\Delta_p(r)$ is continuously differentiable on
$\R^d\times[0,1/2)$.  
In particular, $G_{T,\dm}$ is continuously differentiable on this domain.
Differentiating the maximized expression with respect
to $r$ gives
\[
 \partial_rG_{T,\dm}(p,r)=\Delta_p(r).
\]
Since $G_{T,\dm}(p,0)=0$, this proves \eqref{eq:G-integral}.  
To obtain \eqref{eq:G-derivative}, differentiating
$f_p(\Delta_p(r))=r$ at $r=0$ and using
$h_{T,\dm}(p,0)=K_{T,\dm}(p)^{-1}$ gives
\[
 \partial_r^2G_{T,\dm}(p,0)=\Delta_p'(0)
 =\frac1{f_p'(0)}=2K_{T,\dm}(p).
\]

At $|z|=1/2$, the constraints
\[
 |z|^2\leq\min\{\gamma_+(1-\gamma_-),\gamma_-(1-\gamma_+)\}
\]
force $\gamma_+=\gamma_-=1/2$.  Thus the matrix in
\eqref{eq:endpoint-state}, with $\alpha(p)=z$, is the unique admissible matrix
with pairing $z$.  As $|z|\uparrow1/2$, the maximizing gap tends to infinity,
and \eqref{eq:gamma-plus-explicit}--\eqref{eq:gamma-minus-explicit} show that
both diagonal entries tend to $1/2$.  In particular, $\Gamma^{[z]}(p)$,
$\ell_{T,p}(\Gamma^{[z]}(p))$ and thus also $G_{T,\dm}(p,z)$ are continuous in $z$.
Taking $r\uparrow1/2$ in \eqref{eq:G-dual-lower-bound} proves the dual formula
also at the endpoint, since the expressions inside the supremum increase
with $r$.  The integral formula follows there by monotone convergence.
Nonnegativity and convexity of $G_{T,\dm}(p,\cdot)$ now follow from
\eqref{eq:G-def}.

\emph{Step 3: admissibility of the optimizer.}
Continuity of the pointwise optimizer gives measurability.  It remains to
verify the integrability required by the domain $\mathcal D$ in
\eqref{eq:domain}. 
Let $\alpha\in\Da$ with $|\alpha|\leq1/2$ almost everywhere, and set
\[
 A:=\{p\in\R^d:|\alpha(p)|=1/2\}.
\]
For $p\in A^c$ we proceed as follows.
Choose $R>0$ so large that, for $|p|\geq R$, one has $\xi(p)>\dm$ and
\[
 \tanh\!\left(\frac{\xi(p)+\dm}{2T}\right)
 +\tanh\!\left(\frac{\xi(p)-\dm}{2T}\right)\geq1.
\]
Writing $\Delta=\Delta_p(\alpha(p))$, $D=|\Delta|$, and
\[
 b_D:=\tanh\!\left(\frac{E_D+\dm}{2T}\right)
       +\tanh\!\left(\frac{E_D-\dm}{2T}\right),
\]
the explicit diagonal entries
\eqref{eq:gamma-plus-explicit}--\eqref{eq:gamma-minus-explicit} give
\[
 \gamma_{+,\Delta}+\gamma_{-,\Delta}
 =1-\frac{\xi b_D}{2E_D}
 =1-\frac{b_D}{2}+\frac{b_D}{2}
      \left(1-\frac{\xi}{E_D}\right).
\]
For $|p|\geq R$, one has
$0\leq1-b_D/2\leq C e^{-\xi(p)/T}$,
$1-\xi/E_D\leq D^2/E_D^2$, and, by
\eqref{eq:pairing-fixed-gap}, $D^2/E_D^2=16|\alpha(p)|^2/b_D^2\leq16|\alpha(p)|^2$ because
$b_D\geq1$.  Hence, for $p\in A^c$ with $|p|\geq R$,
\[
 0\leq\gamma_{+,\Delta}(p)+\gamma_{-,\Delta}(p)
 \leq C e^{-\xi(p)/T}+C|\alpha(p)|^2.
\]
This bound also holds for $p\in A$ with $|p|\geq R$ after increasing $C$.
On $|p|<R$ the diagonal densities are bounded by $1$.  Since
$\alpha\in L^2(\R^d,(1+p^2)\dd p)$, the last estimate proves that
$\gamma_{\pm,\Delta}\in L^1(\R^d,(1+p^2)\dd p)$.  Hence
$\Gamma^{[\alpha]}$ is admissible and satisfies the asserted pointwise
identity.  
\end{proof}
\begin{rem}\label{rmk:topology}
    The end of the proof also shows that for each such $\alpha$, pointwise continuity of the optimizer and
the same tail estimate, uniformly for $|t|\leq1$, give
\[
 \lVert \Gamma^{[t\alpha]}-\Gamma_0\rVert_{\rm HS}^2=\int_{\R^d}\Tr_{\C^4}|\Gamma^{[t\alpha]}(p)-\Gamma_0(p)|^2\dd p
 \longrightarrow0\qquad(t\in\R,\ t\to0)
\]
by dominated convergence.
\end{rem}

The following Lemma gives a lower bound on $G_{T,\dm}$, which approximates the quadratic term in the expansion of $G_{T,\dm}$ for small $z$ to arbitrary precision. This will be useful to prove local stability of the normal state.
\begin{lemma}[Uniform small-amplitude bound]\label{lem:small-amplitude}
Let $T>0,\dm \in \BR$.
For every $\eta\in(0,1)$ there is $\rho>0$ such that
\begin{equation}\label{eq:small-amplitude}
 G_{T,\dm}(p,z)\geq(1-\eta)K_{T,\dm}(p)|z|^2
\end{equation}
for all $p$ and $|z|\leq\rho$.  In addition, there is $R<\infty$ such that
the same estimate holds for $|p|\geq R$ and every $|z|\leq1/2$.
\end{lemma}

\begin{proof}
\emph{Roadmap.}  We first obtain an all-amplitude estimate at large momentum,
then use uniform continuity on the remaining compact momentum
set for small amplitudes.

Write $x=|\xi(p)|$.  For $s\geq0$, the numerator in
\eqref{eq:h-response} is at most $2$, so
$h_{T,\dm}(p,s)^{-1}\geq\sqrt{x^2+s}\geq x$.  Therefore
\begin{align}
 \frac{1}{K_{T,\dm}(p)h_{T,\dm}(p,s)}
 &\geq\frac{x}{K_{T,\dm}(p)}\notag\\
 &=\frac12\left[
 \tanh\!\left(\frac{x+\dm}{2T}\right)
 +\tanh\!\left(\frac{x-\dm}{2T}\right)\right]\notag\\
 &\geq1-2e^{-(x-|\dm|)/T}.
 \label{eq:large-p-K-comparison}
\end{align}
Choose $R$ so that the last error is at most $\eta$ for $|p|\geq R$.
Then
\[
 h_{T,\dm}(p,s)\leq\frac1{(1-\eta)K_{T,\dm}(p)}
 \qquad(|p|\geq R,\ s\geq0).
\]
In \eqref{eq:G-def}, test with
$D=2(1-\eta)K_{T,\dm}(p)|z|$.  This gives
\[
 G_{T,\dm}(p,z)
 \geq D|z|-\frac{D^2}{4(1-\eta)K_{T,\dm}(p)}
 =(1-\eta)K_{T,\dm}(p)|z|^2.
\]

On $|p|\leq R$, the function
$K_{T,\dm}(p)h_{T,\dm}(p,s)$ is jointly continuous and equals $1$ at $s=0$.
Hence there is $s_0>0$ such that the same upper bound on $h$ holds for
$0\leq s\leq s_0$ and $|p|\leq R$.  Since $K_{T,\dm}$ is bounded on this
compact set, we can choose $0<\rho<1/2$ so that the preceding test gap
satisfies $D^2\leq s_0$ whenever $|z|\leq\rho$.  This proves the assertion.
\end{proof}

The following lemma provides a lower bound on $G_{T,\dm}$ which is particularly useful to handle large $z$, where the previous bound fails.
\begin{lemma}[Global pointwise coercivity]\label{lem:global-coercivity}
For fixed $T>0$ and $\dm\in \BR$, there is $c_*>0$ such that
\begin{equation}\label{eq:global-G-coercivity}
 G_{T,\dm}(p,z)\geq c_*K_{T,\dm}(p)|z|^2
\end{equation}
for every $p$ and every $|z|\leq1/2$.
\end{lemma}

\begin{proof}

Apply Lemma~\ref{lem:small-amplitude} with $\eta=1/2$.  It remains only to
consider a compact momentum region $|p|\leq R$ and amplitudes
$\rho_0\leq |z|\leq1/2$.  Let
\[
 M=\sup\{h_{T,\dm}(p,s):|p|\leq R,\ 0\leq s\leq1\}<\infty
\]
and choose $D_0=\min\{1,2\rho_0/M\}$.  Testing \eqref{eq:G-def} with $D_0$
gives
\[
 G_{T,\dm}(p,z)\geq D_0|z|-\frac14MD_0^2
 \geq\frac12D_0\rho_0.
\]
Since $K_{T,\dm}$ is bounded above on $|p|\leq R$ and $|z|^2\leq1/4$, this
positive lower bound implies \eqref{eq:global-G-coercivity} on the remaining
compact region.
\end{proof}

For the next proposition recall the definition of $L_{T,\dm}$ from \eqref{eq:interaction-form-definition}.
The next proposition combines Lemmas~\ref{lem:small-amplitude} and
\ref{lem:global-coercivity} with compactness to prove local stability below $\llin$.
\begin{prop}[$L^2$ nonlinear coercivity below the linear threshold]
\label{prop:L2-local-stability}
Fix $T>0$ and $\dm\in \BR$.  Suppose that $L_{T,\dm}$ is compact and
self-adjoint, let $b=\max\spec L_{T,\dm}>0$, and fix
$0\leq\bar\lambda<b^{-1}$.  Then there are
$\rho,c_{\mathrm{loc}}>0$ such that, for every
$0\leq\lambda\leq\bar\lambda$, every $\alpha\in\Da$
with $|\alpha|\leq1/2$ almost everywhere and $\|\alpha\|_2<\rho$ satisfies
\begin{equation}\label{eq:L2-reduced-coercivity}
 \mathcal{R}_\lambda(\alpha)=\int_{\R^d}G_{T,\dm}(p,\alpha(p))\dd p
 -\lambda\langle \alpha,\Vop \alpha\rangle
 \geq c_{\mathrm{loc}}\langle \alpha,K_{T,\dm}\alpha\rangle.
\end{equation}
\end{prop}

\begin{proof}
\emph{Roadmap.}  We argue by contradiction, normalize in the $K$-energy, and
split into small- and large-amplitude regions.  The latter region has vanishing
measure, so compactness of $L$ removes its interaction contribution, while the
pointwise bounds control both pieces.

Abbreviate $K=K_{T,\dm}$, $L=L_{T,\dm}$, and set
$c_0=1-\bar\lambda b>0$.  The spectral theorem gives, uniformly for
$0\leq\lambda\leq\bar\lambda$,
\begin{equation}\label{eq:I-minus-lambda-L}
 \langle u,(1-\lambda L)u\rangle\geq c_0\|u\|_2^2.
\end{equation}

Suppose the proposition is false.  Then there are
$\lambda_n\in[0,\bar\lambda]$ and nonzero $\alpha_n\in\Da$ with
$|\alpha_n|\leq1/2$ almost everywhere such that
$\|\alpha_n\|_2\to0$ and
\[
 \mathcal R_{\lambda_n}(\alpha_n)<\frac1n\langle \alpha_n,K\alpha_n\rangle,
\]
Set $t_n^2=\langle \alpha_n,K\alpha_n\rangle$ and
$u_n=t_n^{-1}K^{1/2}\alpha_n$, so that $\|u_n\|_2=1$ and
\begin{equation}\label{eq:interaction-normalized}
 \langle \alpha_n,\Vop \alpha_n\rangle
 =t_n^2\langle u_n,Lu_n\rangle.
\end{equation}

Choose $\eta\in(0,c_0)$ and let $\rho_0$ be given by
Lemma~\ref{lem:small-amplitude}.  Define
\[
 A_n=\{|\alpha_n|\leq\rho_0\},\qquad B_n=\R^d\setminus A_n,
 \qquad v_n=\1_{A_n}u_n,\quad w_n=\1_{B_n}u_n.
\]
Since $|B_n|\leq\rho_0^{-2}\|\alpha_n\|_2^2\to0$ and $\|w_n\|_2\leq1$,
for every $g\in L^2(\R^d)$ one has
$|\langle g,w_n\rangle|\leq\|\1_{B_n}g\|_2\to0$.  Thus
$w_n\rightharpoonup0$ in $L^2(\R^d)$.  Compactness of $L$ implies
$Lw_n\to0$ strongly, and hence
\begin{equation}\label{eq:interaction-splitting}
 \langle u_n,Lu_n\rangle=\langle v_n,Lv_n\rangle+o(1).
\end{equation}
The small-amplitude and global estimates give
\begin{align*}
 \frac{\mathcal R_{\lambda_n}(\alpha_n)}{t_n^2}
 &\geq(1-\eta)\|v_n\|_2^2+c_*\|w_n\|_2^2
 -\lambda_n\langle u_n,Lu_n\rangle\\
 &=\langle v_n,(1-\lambda_n L)v_n\rangle
 -\eta\|v_n\|_2^2+c_*\|w_n\|_2^2+o(1)\\
 &\geq(c_0-\eta)\|v_n\|_2^2+c_*\|w_n\|_2^2+o(1).
\end{align*}
Since $v_n$ and $w_n$ have disjoint supports and their squared norms sum to
one, the right side has a strictly positive lower limit.  This contradicts
the choice of $\alpha_n$.
\end{proof}

\subsection{The variational first-order mechanism}

The next proposition proves the coexistence and energy-kink statements of the
main theorem, assuming that a trial state has strictly negative free-energy
difference at coupling $\llin$.  The later sections verify this assumption
for the class of interactions in Theorem~\ref{thm:main}.

\begin{prop}[Variational first-order principle]
\label{prop:variational-first-order}
Fix $T>0$ and $\dm>0$.  Suppose that $L_{T,\dm}$ is compact and
self-adjoint, that $b:=\max\spec L_{T,\dm}>0$, and that for
$\llin=b^{-1}$ there is
$\alpha_0\in\Da$, $|\alpha_0|\leq1/2$, with
\begin{equation}\label{eq:abstract-negative-trial}
 \mathcal R_{T,\dm,\llin}(\alpha_0)<0.
\end{equation}
Then
\begin{equation}\label{eq:abstract-threshold-order}
 0<\lfo(T,\dm)
 <\llin(T,\dm)<\infty
\end{equation}
and the BCS functional has a global minimizer for every $\lambda\geq0$.  Moreover:
\begin{enumerate}[label=\textup{(\roman*)}]
\item If $0\leq\lambda<\lfo(T,\dm)$, then $\Gamma_0$ is
the unique global minimizer.
\item At $\lambda=\lfo(T,\dm)$, the normal state and a
paired state $\Gamma^{[\alpha_*]}$, for some $\alpha_*\in\Da$, coexist as distinct global minimizers, with
\begin{equation}\label{eq:coexistence-main-2}
 \alpha_*\neq0,\qquad \langle \alpha_*,\Vop \alpha_*\rangle>0,\qquad
 \int_{\R^d}G_{T,\dm}(p,\alpha_*(p))\dd p
 =\lfo(T,\dm)\langle \alpha_*,\Vop \alpha_*\rangle.
\end{equation}
\item If
$\lfo(T,\dm)<\lambda<\llin(T,\dm)$,
then every global minimizer is paired, while $\Gamma_0$ remains a strict local minimizer, that is, for each
$0\leq\bar\lambda<\llin(T,\dm)$ there are
$\rho,c_{\mathrm{loc}}>0$ such that, uniformly for
$0\leq\lambda\leq\bar\lambda$,
\begin{equation}\label{eq:L2-local-coercivity-main-2}
 \mathcal E_{T,\dm,\lambda}(\Gamma)
 \geq c_{\mathrm{loc}}\langle \alpha,K_{T,\dm}\alpha\rangle
\end{equation}
whenever $\|\alpha\|_2<\rho$.
In particular, every paired minimizer with nonpositive energy and
$0\leq\lambda\leq\bar\lambda$ satisfies $\|\alpha\|_2\geq\rho$.
Choosing $\bar\lambda>\lfo$ shows that the pairing density jumps at $\lfo$ in $L^2$.
\item Finally,
with
\begin{equation}\label{eq:minimized-energy-main-2}
 e_{T,\dm}(\lambda):=\inf_{\Gamma\in\mathcal D}
 \mathcal E_{T,\dm,\lambda}(\Gamma),
\end{equation}
one has $e_{T,\dm}(\lambda)=0$ for
$0\leq\lambda\leq\lfo(T,\dm)$ and, for $h>0$,
\begin{equation}\label{eq:kink-main-2}
 e_{T,\dm}(\lfo+h)
 \leq-h\langle \alpha_*,\Vop \alpha_*\rangle<0.
\end{equation}
The function $e_{T,\dm}$ is concave, and its one-sided derivatives at
coexistence satisfy
\begin{equation}\label{eq:one-sided-slopes-main-2}
 e'_{T,\dm,-}(\lfo)=0,
 \qquad
 e'_{T,\dm,+}(\lfo)
 \leq-\langle \alpha_*,\Vop \alpha_*\rangle<0.
\end{equation}
Thus the minimized energy has a first-order kink.
\end{enumerate}

\end{prop}

\begin{proof}
For brevity in the proof we drop the indices $T,\dm$.  Every pairing density below is understood
to belong to $\Da$ and to satisfy $|\alpha|\leq1/2$ almost
everywhere.

Existence in $d=3$ is stated in \cite[Proposition~1]{FreijiHainzlSeiringer2012},
using the direct method of \cite[Proposition~2]{HainzlHamzaSeiringerSolovej2008}.
The same argument extends to $d=1,2$ under \eqref{eq:V-assumptions}, as noted
for the balanced functional in \cite[Proposition~1.2]{HenheikLauritsenRoos2024}.  Lemma~\ref{lem:fixed-gap-duality}
identifies the pairing densities of global minimizers with minimizers of
$\mathcal R_\lambda$.

We first order the thresholds and establish stability away from coexistence.
We then take a limit of paired minimizers to obtain coexistence; compactness
and local coercivity prevent the limit from vanishing.  Finally, that
coexistence state gives the claimed energy kink.

\emph{Positivity of $\lfo$.}
Lemma~\ref{lem:global-coercivity} gives
$\int G(p,\alpha(p))\dd p\geq c_*\langle\alpha,K\alpha\rangle$
for every admissible pairing density.  The spectral bound on $L$ gives
$\langle\alpha,\Vop\alpha\rangle\leq b\langle\alpha,K\alpha\rangle$.
Taking the infimum in \eqref{eq:lambda-fo} yields $\lfo\geq c_*/b>0$.

\emph{Separation of the thresholds.}
Since $G\geq0$, the assumption $\mathcal R_{\llin}(\alpha_0)<0$ implies
$\langle\alpha_0,\Vop\alpha_0\rangle>0$ and
\[
 \lfo\leq
 \frac{\int G(p,\alpha_0(p))\dd p}{\langle\alpha_0,\Vop\alpha_0\rangle}
 <\llin.
\]

\emph{Case $0\leq \lambda<\lfo$.}
We want to show that $\Gamma_0$ is the unique minimizer.
Let $\alpha\neq0$.  There are two
cases.  If $\langle \alpha,\Vop \alpha\rangle>0$, then
\begin{align*}
 \mathcal R_{\lambda}(\alpha)
 &=\langle \alpha,\Vop \alpha\rangle
 \left(
  \frac{\displaystyle\int G(p,\alpha(p))\dd p}
       {\langle \alpha,\Vop \alpha\rangle}-\lambda
 \right)\\
 &\geq
 \langle \alpha,\Vop \alpha\rangle(\lfo-\lambda)>0.
\end{align*}
If $\langle \alpha,\Vop \alpha\rangle\leq0$, then by Lemma~\ref{lem:global-coercivity}
\[
 \mathcal R_{\lambda}(\alpha)
 \geq\int G(p,\alpha(p))\dd p
 \geq c_*\langle \alpha,K\alpha\rangle>0.
\]
Thus every state with $\alpha\neq0$ has positive free-energy difference.
If $\alpha=0$, equality in \eqref{eq:relative-entropy-identity}, with
$\Delta=0$, holds only for $\Gamma=\Gamma_0$.  This proves uniqueness.

\emph{Case $\lfo<\lambda <\llin$.}
The bound \eqref{eq:L2-local-coercivity-main-2} follows from
Proposition~\ref{prop:L2-local-stability} and the constrained definition of $G$.
Together with the preceding uniqueness argument at $\alpha=0$, it gives
strict local minimality.  It also yields the asserted uniform $L^2$ lower
bound for every paired minimizer with nonpositive energy.
To prove global instability, choose an admissible $\alpha_0$ with
$\langle\alpha_0,\Vop\alpha_0\rangle>0$ and 
\begin{equation}
    \frac{\int G(p,\alpha_0(p))\dd p}{\langle \alpha_0,\Vop \alpha_0\rangle} < \lambda.
\end{equation}
(Such an $\alpha_0$ exists by the definition of $\lfo$ and since $\lfo<\lambda$.)
By Lemma~\ref{lem:fixed-gap-duality} we have
\begin{equation}
 \mathcal E_{\lambda}(\Gamma^{[\alpha_0]})
 = \int_{\R^d}G(p,\alpha_0(p))\dd p
-\lambda \langle \alpha_0,\Vop \alpha_0\rangle 
=\langle \alpha_0,\Vop \alpha_0\rangle\left(\frac{\int G(p,\alpha_0(p))\dd p}{\langle \alpha_0,\Vop \alpha_0\rangle}
-\lambda  \right)<0.
\end{equation}

\emph{Case $\lambda=\lfo$.} 
Fix
\begin{equation}\label{eq:bar-lambda-choice}
 \lfo<\bar\lambda<\llin
\end{equation}
and choose $\lambda_n\downarrow\lfo$ with $\lambda_n>\lfo$ and
$\lambda_n\leq\bar\lambda$.  Let $\alpha_n\in\Da$ be the pairing density
of a global minimizer at $\lambda_n$, and set
$u_n=K^{1/2}\alpha_n\in L^2(\R^d)$.  Then
\begin{equation}\label{eq:negative-minimizers}
 \mathcal R_{\lambda_n}(\alpha_n)<0,
 \qquad \langle \alpha_n,\Vop \alpha_n\rangle>0.
\end{equation}
We first show that the sequence $(u_n)$ is bounded in $L^2$.
Choose $\eta>0$ so small that
\begin{equation}\label{eq:eta-bar-lambda}
 1-\eta-\bar\lambda b>0.
\end{equation}
By Lemma~\ref{lem:small-amplitude}, there is $R<\infty$ such that
\[
 G(p,z)\geq(1-\eta)K(p)|z|^2
 \qquad(|p|\geq R,\ |z|\leq1/2).
\]
The pointwise bound $|\alpha_n|\leq1/2$ gives
\begin{equation}\label{eq:low-momentum-bound}
 \|\1_{\{|p|<R\}}u_n\|_2^2
 =\int_{|p|<R}K(p)|\alpha_n(p)|^2\dd p
 \leq\frac14\int_{|p|<R}K(p)\dd p\leq C<\infty.
\end{equation}
Using $G\geq0$ on the low-momentum region and $\langle \alpha_n,\Vop \alpha_n\rangle=\langle u_n,Lu_n\rangle
 \leq b\|u_n\|_2^2$, we obtain
\begin{align}
 0&>\mathcal R_{T,\dm,\lambda_n}(\alpha_n)\notag\\
 &\geq(1-\eta)\|\1_{\{|p|\geq R\}}u_n\|_2^2
       -\bar\lambda b\|u_n\|_2^2\notag\\
 &\geq(1-\eta-\bar\lambda b)
       \|\1_{\{|p|\geq R\}}u_n\|_2^2
       -C.
 \label{eq:uniform-tail-bound}
\end{align}
Equations \eqref{eq:eta-bar-lambda} and \eqref{eq:uniform-tail-bound}, together
with \eqref{eq:low-momentum-bound}, prove that $(u_n)$ is bounded.

By the Banach--Alaoglu theorem, after passing to a subsequence there is a $u_*\in L^2$ with $u_n\rightharpoonup u_*$.
The constraint on $\alpha$ is equivalent to the constraint $|u_n|\leq \frac{1}{2}K^{1/2}$.
The set of $L^2$ functions satisfying this constraint and the norm bound on $(u_n)$ is convex and closed in $L^2$. The Hahn--Banach theorem implies that the set is weakly closed, hence $u_*$ satisfies the constraints and $\alpha_*=K^{-1/2}u_*$ is admissible.

Compactness of $L$ gives
\begin{equation}\label{eq:interaction-convergence-coexistence}
 \langle \alpha_n,\Vop \alpha_n\rangle
 =\langle u_n,Lu_n\rangle
 \longrightarrow\langle u_*,Lu_*\rangle
 =\langle \alpha_*,\Vop \alpha_*\rangle.
\end{equation}
In particular, the interaction energies on the left are uniformly bounded.
Recall that
\begin{equation}\label{eq:Rfo-nonnegative}
 \mathcal R_{\lfo}(\alpha)\geq0
 \quad\text{for every admissible }\alpha.
\end{equation} 
Therefore
\begin{align}
 0&\leq\mathcal R_{\lfo}(\alpha_n)\notag\\
 &=\mathcal R_{\lambda_n}(\alpha_n)
   +(\lambda_n-\lfo)
     \langle \alpha_n,\Vop \alpha_n\rangle\notag\\
 &<(\lambda_n-\lfo)
     \langle \alpha_n,\Vop \alpha_n\rangle\longrightarrow0.
 \label{eq:Rfo-convergence}
\end{align}
The functional
\[
 u\longmapsto\int_{\R^d}G\bigl(p,K(p)^{-1/2}u(p)\bigr)\dd p,
 \qquad u\in L^2(\R^d),
\]
is convex and norm lower semicontinuous, by Lemma~\ref{lem:fixed-gap-duality}
and Fatou's lemma; here $G(p,z)=+\infty$ for $|z|>1/2$.  It is therefore
weakly lower semicontinuous.  Together with
\eqref{eq:interaction-convergence-coexistence}, this gives
\[
 \mathcal R_{\lfo}(\alpha_*)
 \leq\liminf_{n\to\infty}
 \mathcal R_{\lfo}(\alpha_n)=0.
\]
Combining this with \eqref{eq:Rfo-nonnegative} yields
\(0=\mathcal R_{\lfo}(\alpha_*)=\int G(p,\alpha_*(p))\dd p-\lfo \langle\alpha_*, \Vop \alpha_*\rangle.\)

It remains to exclude $\alpha_*=0$.  Suppose that $u_*=0$.  Then
\eqref{eq:interaction-convergence-coexistence} gives
$\langle \alpha_n,\Vop \alpha_n\rangle\to0$.  From
\eqref{eq:negative-minimizers},
\[
 0\leq\int G(p,\alpha_n(p))\dd p
 <\lambda_n\langle \alpha_n,\Vop \alpha_n\rangle\longrightarrow0.
\]
Lemma~\ref{lem:global-coercivity} now implies
$u_n\to0$ strongly.  Since $K^{-1/2}$ is bounded, also $\alpha_n\to0$ in $L^2$.
Apply Proposition~\ref{prop:L2-local-stability} with the fixed number
$\bar\lambda$ from \eqref{eq:bar-lambda-choice}.  For all sufficiently large
$n$ it gives
\[
 \mathcal R_{\lambda}(\alpha_n)
 \geq c_{\mathrm{loc}}\|u_n\|_2^2>0
\]
for all $0\leq\lambda\leq\bar\lambda$, in particular for $\lambda=\lambda_n$.
This contradicts \eqref{eq:negative-minimizers}.  Hence $\alpha_*\neq0$.  If
$\langle \alpha_*,\Vop \alpha_*\rangle\leq0$,  by Lemma~\ref{lem:global-coercivity} 
$ \mathcal R_{\lfo}(\alpha_*)$ would be strictly positive.  
Therefore
$\langle \alpha_*,\Vop \alpha_*\rangle>0$, and the claim \eqref{eq:coexistence-main-2} follows.

\emph{Energy kink.}
Use $\alpha_*$ as a trial state.  For every $h>0$,
\begin{equation}
 e_{T,\dm}(\lfo+h)
 \leq\mathcal R_{T,\dm,\lfo+h}(\alpha_*)
 =\mathcal R_{T,\dm,\lfo}(\alpha_*)
   -h\langle \alpha_*,\Vop \alpha_*\rangle
 =-h\langle \alpha_*,\Vop \alpha_*\rangle<0.
 \label{eq:kink-trial-detailed}
\end{equation}
For $0\leq \lambda\leq \lfo$ the normal state is a global minimizer and $e_{T,\dm}(\lambda)=0$.
For each admissible state the energy is affine in $\lambda$; hence their
infimum $e_{T,\dm}$ is concave. Its left derivative at
coexistence is therefore zero.  
From \eqref{eq:kink-trial-detailed} we obtain
\[
 \frac{e_{T,\dm}(\lfo+h)
       -e_{T,\dm}(\lfo)}{h}
 \leq-\langle \alpha_*,\Vop \alpha_*\rangle.
\]
Letting $h\downarrow0$ and using existence of the right derivative of a finite
concave function proves the asserted strict slope inequality.
\end{proof}

\section{The linear threshold: Birman--Schwinger analysis}
\label{sec:BS}

In this section we discuss compactness and the low $T,\dm$ asymptotics of the Birman-Schwinger operators.

\begin{lemma}[Form bound and compactness]\label{lem:compact-sandwich}
Under the integrability assumptions on $V$ in \eqref{eq:V-assumptions}, the operator
\[
 \Wop(1-\Delta)^{-1/2}:L^2(\R^d)\longrightarrow L^2(\R^d)
\]
is compact.
Consequently, if $g:\R^d\to\C$ is measurable and satisfies
$|g(p)|\leq C(1+p^2)^{-1}$ almost everywhere, then
$\Wop g(-i\nabla)\Wop$ extends to a compact operator on $L^2(\R^d)$.
\end{lemma}

\begin{proof}
\emph{Roadmap.}  Sobolev form bounds give boundedness, approximation by bounded
compactly supported potentials gives compactness, and a simple factorization
then treats every decaying multiplier $g$.

For $u\in H^1(\R^d)$, H\"older's inequality and Sobolev embedding give
\begin{equation}\label{eq:V-form-bound}
 \int |V||u|^2
 \leq C_V\|u\|_{H^1}^2.
\end{equation}
Indeed, one uses $H^1(\R)\hookrightarrow L^\infty(\R)$ in $d=1$,
$H^1(\R^2)\hookrightarrow L^q(\R^2)$ for every finite $q$ in $d=2$, with
$q=2(1+\varepsilon)/\varepsilon$, and
$H^1(\R^3)\hookrightarrow L^6(\R^3)$ in $d=3$.  Thus
$\Wop (1-\Delta)^{-1/2}$ is bounded.  Approximate $|V|$ in the relevant
$L^p$ norm by nonnegative bounded compactly supported functions.  For such
functions, compactness follows from the Rellich theorem
\cite[Theorem 6.3]{AdamsFournier}.  Applying \eqref{eq:V-form-bound} to the
squared differences of their square roots shows convergence of the corresponding
operators in norm and therefore gives compactness for $V$.

Writing
\[
 g(-i\nabla)=(1-\Delta)^{-1/2}h(-i\nabla)(1-\Delta)^{-1/2},
 \qquad h(p)=(1+p^2)g(p),
\]
gives
\[
 \Wop g(-i\nabla)\Wop
 =\bigl(\Wop(1-\Delta)^{-1/2}\bigr)h(-i\nabla)
 \bigl(\Wop(1-\Delta)^{-1/2}\bigr)^*,
\]
which is compact because the outer factors are compact and the middle factor is bounded.
\end{proof}

For $\dm>0$, the zero-temperature multiplier needed below is
\begin{equation}\label{eq:m-zero}
K_{0,\dm}^{-1}(p)=
 \frac{\1_{\{|p^2-\mub|>\dm\}}}{|p^2-\mub|}.
\end{equation}
Define on $L^2(\R^d)$
\begin{equation}\label{eq:B-A-def}
 B_{T,\dm}=\Wop K_{T,\dm}^{-1}(-i\nabla)\Wop,
 \qquad A_{T,\dm}=JB_{T,\dm}.
\end{equation}
For fixed $\dm>0$, Lemma~\ref{lem:compact-sandwich} implies that
$B_{T,\dm}$ is nonnegative and compact for $T\geq0$.  Define the bounded extensions
$X_{T,\dm}:L^2(\R^d)\to L^2(\R^d)$ and
$Y_{T,\dm}:L^2(\R^d)\to L^2(\R^d)$ by
\[
 X_{T,\dm}=K_{T,\dm}^{-1/2}\F \Wop,
 \qquad Y_{T,\dm}=J \Wop \F^{-1}K_{T,\dm}^{-1/2}.
\]
At $T=0$, set $L_{0,\dm}:=X_{0,\dm}Y_{0,\dm}$.
Then $L_{T,\dm}=X_{T,\dm}Y_{T,\dm}$ and
$A_{T,\dm}=Y_{T,\dm}X_{T,\dm}$ for $T\geq0$.  Hence their nonzero eigenvalues, including
algebraic multiplicities, agree.  In particular, although $A_{T,\dm}$ need
not be self-adjoint, its nonzero spectrum is real because $L_{T,\dm}$ is
self-adjoint.

The following result is based on the standard Fermi-shell decomposition, which was first used in the spin-balanced case in \cite{FHNS2007,HainzlSeiringer2008}.
In the spin-polarized case, \cite[Theorems 2--4]{FreijiHainzlSeiringer2012}
derive spectral and eigenvector asymptotics for related Birman--Schwinger
operators as $(T,\dm)\to(0,0)$.
We include the proof to establish the uniform convergence needed below,
with the limits taken first as $T\downarrow0$ and then as $\dm\downarrow0$.

Recall the definition of $\mathcal{T}_s$ from \eqref{eq:T-s-def}.
\begin{prop}[Norm limits and critical eigenvectors]\label{prop:BS}
Let $\mub>0$.
Under the assumptions on $V$ in \eqref{eq:V-assumptions}, the following statements hold.
\begin{enumerate}[label=\textup{(\alph*)}]
\item For every fixed $\dm>0$,
\begin{equation}\label{eq:A-T-limit}
 \|A_{T,\dm}-A_{0,\dm}\|_{\op}\longrightarrow0
 \qquad(T\downarrow0).
\end{equation}
\item With $\ell_{\dm}=\ln(\mub/\dm)$,
\begin{equation}\label{eq:A-delta-limit}
 \left\|\ell_{\dm}^{-1}A_{0,\dm}
 -\mub^{(d-2)/2}J\mathcal T_{\mub}^*\mathcal T_{\mub}
 \right\|_{\op}\longrightarrow0
 \qquad(\dm\downarrow0).
\end{equation}
\item If $\nu_{\mub}=\sup \sigma(\Vop_{\mub})>0$ is a simple eigenvalue, 
then
for all sufficiently small $\dm$ and subsequently sufficiently small $T$,
$b_{T,\dm}=\max\spec L_{T,\dm}$ is positive and simple.  There are normalized
eigenvectors $\psi_{T,\dm}\in L^2(\R^d)$ satisfying
\begin{equation}\label{eq:A-eigen}
 A_{T,\dm}\psi_{T,\dm}=b_{T,\dm}\psi_{T,\dm},
 \qquad \|\psi_{T,\dm}\|_2=1,
\end{equation}
whose phases can be chosen so that
\begin{align}
 \psi_{T,\dm}&\longrightarrow\psi_{0,\dm}
 &&\text{in }L^2\quad(T\downarrow0),\label{eq:psi-T}\\
 \psi_{0,\dm}&\longrightarrow\psi_*
 &&\text{in }L^2\quad(\dm\downarrow0),\label{eq:psi-delta}
\end{align}
where
\begin{equation}\label{eq:A-star-eigen}
 \mub^{(d-2)/2}J\mathcal T_{\mub}^*\mathcal T_{\mub}\psi_*
 =\nu_{\mub}\psi_*.
\end{equation}
Moreover, the functions
\begin{equation}\label{eq:q-eigen-def}
 q_{T,\dm}=\F\Wop\psi_{T,\dm}\quad(T\geq0),
 \qquad q_*=\F\Wop\psi_*,
\end{equation}
are bounded continuous functions from $\R^d$ to $\C$.
Moreover, $q_{T,\dm}\to q_{0,\dm}$ uniformly on $\R^d$ as $T\downarrow0$,
$q_{0,\dm}\to q_*$ uniformly on $\R^d$ as $\dm\downarrow0$, and
\begin{equation}\label{eq:q-star-nonzero}
 \int_{\Sph^{d-1}}|q_*(\sqrt{\mub}\,\omega)|^4\dd\omega>0.
\end{equation}
\end{enumerate}
\end{prop}

\begin{proof}
\emph{Roadmap.}  We first pass $T\downarrow0$ in operator norm, then isolate
the logarithmic Fermi-shell contribution as $\dm\downarrow0$.  Perturbation
theory identifies the simple top eigenvalue, and a final Fourier estimate gives the desired
uniform convergence.

\emph{Step 1: the limit $T\downarrow0$.}
Write $x=|p^2-\mub|$ and
\[
 D_T(x)=\tanh\!\left(\frac{x+\dm}{2T}\right)
 +\tanh\!\left(\frac{x-\dm}{2T}\right),
 \qquad K_{T,\dm}^{-1}=\frac{D_T(x)}{2x}.
\]
Fix $0<\eta<\dm/2$.  On $x\leq\dm-\eta$, the mean-value theorem and
$\operatorname{sech}^2 y\leq4e^{-2|y|}$ give
\[
 0\leq K_{T,\dm}^{-1}(p)\leq \frac{2}{T}e^{-\eta/T}.
\]
On $x\geq\dm+\eta$, the bound $1-\tanh(y)\leq 2e^{-2y}$ for $y\geq0$ gives
\[
 \left|K_{T,\dm}^{-1}(p)-x^{-1}\right|
 \leq 2x^{-1}e^{-(x-\dm)/T}.
\]
On $|x-\dm|<\eta$, both $K_{T,\dm}^{-1}$ and $x^{-1}$ are bounded by $2/\dm$, and the measure
of this shell tends to zero with $\eta$.  These estimates, followed first by
$T\downarrow0$ and then by $\eta\downarrow0$, prove
\begin{equation}\label{eq:m-L1-convergence}
 \|K_{T,\dm}^{-1}-K_{0,\dm}^{-1}\|_{L^1(\R^d)}\longrightarrow0.
\end{equation}
If $g\in L^1(\R^d)$, the kernel of $\Wop g(-i\nabla)\Wop$ is
$(2\pi)^{-d/2}\Wop(x)(\F^{-1}g)(x-y)\Wop(y)$, and hence
\begin{equation}\label{eq:HS-bound}
 \|\Wop g(-i\nabla)\Wop\|_{\op}
 \leq\|\Wop g(-i\nabla)\Wop\|_{\mathrm{HS}}
 \leq(2\pi)^{-d}\|V\|_1\|g\|_1.
\end{equation}
Equations \eqref{eq:m-L1-convergence} and \eqref{eq:HS-bound} prove
\eqref{eq:A-T-limit}.

\emph{Step 2: the logarithmic shell asymptotic.}
We first separate the contribution away from the Fermi sphere.
Choose $0<\varepsilon<\mub/2$ and let $\delta_\mu<\varepsilon$.
By Lemma~\ref{lem:compact-sandwich}, $B_{0,\varepsilon}$ is compact and hence $\ell_{\dm}^{-1} B_{0,\varepsilon}$ vanishes in operator norm as $\dm\to 0$.
For $f\in L^2(\R^d)$, polar coordinates and the substitution $s=|p|^2$ give
\begin{equation}\label{eq:coarea-B}
 \langle f,(B_{0,\dm}-B_{0,\varepsilon})f\rangle
 =\frac12\int_{\varepsilon>|s-\mub|>\dm}
 \frac{s^{(d-2)/2}}{|s-\mub|}
 \|\mathcal T_s f\|_{L^2(\Sph^{d-1})}^2\dd s.
\end{equation}

The map $s\mapsto\mathcal T_s$ is continuous in Hilbert--Schmidt norm.  Indeed,
its squared difference is
\[
 (2\pi)^{-d}\int_{\Sph^{d-1}}\int_{\R^d}
 \left|e^{-i\sqrt{s}\omega\cdot x}
 -e^{-i\sqrt{s_0}\omega\cdot x}\right|^2|V(x)|\dd x\dd\omega,
\]
which tends to zero by dominated convergence.  Consequently,
\begin{equation}\label{eq:C-s-continuity}
 \CC_s:=s^{(d-2)/2}\mathcal T_s^*\mathcal T_s
\end{equation}
defines a compact operator on $L^2(\R^d)$ and is continuous in operator norm near $s=\mub$.
Since
\[
 \frac12\int_{\dm<|s-\mub|<\varepsilon}
 \frac{\dd s}{|s-\mub|}=\ln(\varepsilon/\dm),
\]
it remains to show that
$\frac12\int_{\dm<|s-\mub|<\varepsilon}|s-\mub|^{-1}
(\CC_s-\CC_{\mub})\dd s=o(\ell_{\dm})$ in operator norm.  Given
$\zeta>0$, choose $r<\varepsilon$ so that
$\|\CC_s-\CC_{\mub}\|_{\op}<\zeta$ whenever $|s-\mub|<r$.  The contribution of
this region divided by $\ell_{\dm}$ is at most
$\zeta\ln(r/\dm)/\ell_{\dm}$, while the contribution of
$r\leq|s-\mub|<\varepsilon$ is $O(\ell_{\dm}^{-1})$.  Letting first
$\dm\downarrow0$ and then $\zeta\downarrow0$ proves
\begin{equation}\label{eq:B-delta-limit}
 \left\|\ell_{\dm}^{-1}B_{0,\dm}
 -\mub^{(d-2)/2}\mathcal T_{\mub}^*\mathcal T_{\mub}
 \right\|_{\op}\longrightarrow0.
\end{equation}
Multiplication by the bounded operator $J$ proves
\eqref{eq:A-delta-limit}.  

\emph{Step 3: Convergence of eigenvalues and eigenvectors.}
The nonzero spectra of
\[
 \mub^{(d-2)/2}J\mathcal T_{\mub}^*\mathcal T_{\mub}
 \quad\text{and}\quad
 \Vop_{\mub}=\mub^{(d-2)/2}\mathcal T_{\mub}J\mathcal T_{\mub}^*
\]
agree, including algebraic multiplicities, by the $XY$--$YX$
correspondence.  Thus $\nu_{\mub}$ is a simple isolated eigenvalue of the
first operator.  
Norm convergence and the resolvent identity imply convergence of the corresponding Riesz projections; see \cite{Kato1995}.  
Hence, for small $\dm$, the largest eigenvalue of $\ell_{\dm}^{-1}A_{0,\dm}$ is simple and close to $\nu_{\mub}$, in particular positive.  
By the $XY$--$YX$ correspondence, $b_{0,\dm}=\max\spec L_{0,\dm}$ is
positive and simple, with $\ell_{\dm}^{-1}b_{0,\dm}\to\nu_{\mub}$. 
The same argument using \eqref{eq:A-T-limit} proves the assertion for small $T>0$.
Since the rank-one Riesz projections converge in norm, their
normalized spanning vectors satisfy \eqref{eq:psi-T} and
\eqref{eq:psi-delta} after choosing suitable phases.

\emph{Step 4: uniform convergence and nonvanishing.}
The functions $q_{T,\dm},q_*$ are bounded and continuous because they are Fourier
transforms of $\Wop\psi_{T,\dm},\Wop\psi_*\in L^1(\R^d)$.
Similarly, the two uniform convergences follow from $\Wop\in L^2$ and
\begin{equation}\label{eq:q-uniform-bound}
 \|\F(\Wop(\psi-\varphi))\|_\infty
 \leq(2\pi)^{-d/2}\|\Wop\|_2\|\psi-\varphi\|_2.
\end{equation}
If $q_*(\sqrt{\mub}\omega)$ vanished almost everywhere on $\BS^{d-1}$, then
$\mathcal T_{\mub}\psi_*=0$, and the left side of
\eqref{eq:A-star-eigen} would vanish, contradicting
$\nu_{\mub}>0$.  
Hence \eqref{eq:q-star-nonzero} follows.
\end{proof}

For $T>0$ and $\dm>0$ in the parameter range of Proposition~\ref{prop:BS}(c),
the critical eigenvector produces a zero mode of the Hessian.  Define
\begin{equation}\label{eq:alpha-critical}
 \alpha_{T,\dm}=b_{T,\dm}^{-1}K_{T,\dm}^{-1}\F(\Wop\psi_{T,\dm}).
\end{equation}
Then
\begin{equation}\label{eq:linear-equation}
 K_{T,\dm}\alpha_{T,\dm}
 =\llin(T,\dm)\Vop \alpha_{T,\dm},
 \qquad
 \Vop \alpha_{T,\dm}=q_{T,\dm}.
\end{equation}
Both identities are understood in the form sense; in the second,
$\Vop\alpha_{T,\dm}$ is represented by the bounded continuous function
$q_{T,\dm}$.  Indeed, applying
$J\Wop\F^{-1}$ to \eqref{eq:alpha-critical} and using
\eqref{eq:A-eigen} gives
$J\Wop\F^{-1}\alpha_{T,\dm}=\psi_{T,\dm}$.  The second identity follows, and
the first follows from $\llin(T,\dm)=b_{T,\dm}^{-1}$.  Moreover, $\alpha_{T,\dm}\in\Da$ by \eqref{eq:K-growth}, since
$K_{T,\dm}^{1/2}\alpha_{T,\dm}
=b_{T,\dm}^{-1}X_{T,\dm}\psi_{T,\dm}\in L^2(\R^d)$.  Since
$K_{T,\dm}^{-1}$ and $q_{T,\dm}$ are bounded, $\alpha_{T,\dm}\in L^\infty$ as well.

\section{The nonlinear problem: fourth-order expansion and its sign}
\label{sec:quartic}

The main conclusion of this section is Proposition~\ref{prop:negative-quartic}.
Its proof combines a uniform fourth-order expansion with two successive limits
for the quartic coefficient: first $T\downarrow0$, then $\dm\downarrow0$.
The following lemmas establish these ingredients.

\begin{lemma}[Uniform fourth-order expansion]\label{lem:quartic-expansion}
For fixed $T>0$ and $\dm>0$, there are $\rho\in(0,1/2)$ and $C>0$ such that,
for every $p\in\R^d$ and $z\in\C$ with $|z|\leq\rho$,
\begin{equation}\label{eq:G-expansion}
 G_{T,\dm}(p,z)
 =K_{T,\dm}(p)|z|^2
 -2K_{T,\dm}(p)^4j_{T,\dm}(p)|z|^4
 +R_{T,\dm}(p,|z|),
\end{equation}
where
\begin{equation}\label{eq:j-def}
 \begin{split}
 j_{T,\dm}(p)
 &=-\frac1{2K_{T,\dm}(p)\xi(p)^2}\\
 &\quad+
 \frac{\operatorname{sech}^2\!\left(\frac{|\xi(p)|+\dm}{2T}\right)
 +\operatorname{sech}^2\!\left(\frac{|\xi(p)|-\dm}{2T}\right)}
 {8T\xi(p)^2},
 \end{split}
\end{equation}
the singularity at $\xi=0$ being removable, and
\begin{equation}\label{eq:R-bound}
 |R_{T,\dm}(p,|z|)|\leq C(1+p^2)|z|^6.
\end{equation}
\end{lemma}

\begin{proof}
We expand $G_{T,\dm}$ by integrating the Taylor expansion of $\Delta_p$;
the only estimate requiring a separate argument is a uniform bound on
the fifth derivative of $\Delta_p$.
Since $G_{T,\dm}(p,z)$ depends only on $|z|$, we set $r=|z|$ and use
\eqref{eq:G-integral}:
\[
 G_{T,\dm}(p,z)=\int_0^r\Delta_p(s)\dd s.
\]
Thus, after computing the linear and cubic coefficients of $\Delta_p$,
we will prove that, for some $\rho\in(0,1/2)$,
\begin{equation}\label{eq:Delta-fifth-bound}
 |\Delta_p^{(5)}(s)|\leq C(1+p^2)
 \qquad (p\in\R^d,\ |s|\leq\rho).
\end{equation}
Here and below, $C$ may depend on $T$, $\dm$, and $\mub$, but not on
$p$ or $s$.

\emph{Step 1: the Taylor coefficients.}
The map
\[
 \R^d\times\R\ni(p,D)\longmapsto
 \frac12D h_{T,\dm}(p,D^2)\in\R
\]
is smooth and odd in $D$. At $\xi(p)=D=0$, smoothness follows from
\eqref{eq:h-response}: the numerator is odd analytic in
$\sqrt{\xi(p)^2+D^2}$, so division by this square root leaves a
convergent power series in $\xi(p)^2+D^2$.
The derivative with respect to $D$ at $D=0$ is
$(2K_{T,\dm}(p))^{-1}>0$. The inverse function theorem applied to
\eqref{eq:a-Delta} therefore makes $(p,r)\mapsto\Delta_p(r)\in\R$
smooth near $\R^d\times\{0\}$.
Uniqueness in \eqref{eq:a-Delta} also gives
$\Delta_p(-r)=-\Delta_p(r)$ for $|r|<1/2$.

Direct differentiation of \eqref{eq:h-response} gives
\[
 h_{T,\dm}(p,0)=K_{T,\dm}(p)^{-1},
 \qquad
 \partial_s h_{T,\dm}(p,0)=j_{T,\dm}(p).
\]
The same power-series argument justifies the second identity at
$\xi(p)=0$ and shows that the singularity in \eqref{eq:j-def} is removable.
Since $\Delta_p$ is odd, its even-order derivatives vanish at zero.
Differentiating \eqref{eq:a-Delta} once and three times at $r=0$ gives
\[
 1=\frac{\Delta_p'(0)}{2K_{T,\dm}(p)},
 \qquad
 0=\frac{\Delta_p'''(0)}{2K_{T,\dm}(p)}
   +3j_{T,\dm}(p)\bigl(\Delta_p'(0)\bigr)^3.
\]
Consequently,
\[
 \Delta_p'(0)=2K_{T,\dm}(p),
 \qquad
 \Delta_p'''(0)=-48K_{T,\dm}(p)^4j_{T,\dm}(p).
\]
After integration, these give the quadratic and quartic terms in
\eqref{eq:G-expansion}.

\emph{Step 2: the uniform remainder.}
Choose $R>0$ large enough for Lemma~\ref{lem:rescaled-response} below
to apply. The set $\{p\in\R^d:|\xi(p)|\leq R\}$ is compact,
and $(2K_{T,\dm}(p))^{-1}$ has a positive minimum on it.
A finite collection of the inverse-function neighborhoods from Step~1
covers this set. Shrinking to a common interval in $s$ and using joint
smoothness therefore gives a common
$\rho\in(0,1/2)$ such that
\[
 \sup_{|\xi(p)|\leq R}\sup_{|s|\leq\rho}
 |\Delta_p^{(5)}(s)|\leq C.
\]
After decreasing $\rho$ if necessary, Lemma~\ref{lem:rescaled-response}
gives the complementary estimate
\[
 |\Delta_p^{(5)}(s)|\leq C|\xi(p)|\leq C(1+p^2)
 \qquad (|\xi(p)|\geq R,\ |s|\leq\rho).
\]
Together these prove \eqref{eq:Delta-fifth-bound}.

Now $G_{T,\dm}(p,\cdot)$ is even and smooth near zero, with
$\partial_r^6G_{T,\dm}(p,r)=\Delta_p^{(5)}(r)$.
Taylor's theorem through degree five, with the coefficients from
Step~1, yields \eqref{eq:G-expansion} and the integral remainder
\[
 \begin{split}
 R_{T,\dm}(p,r)
 &=\frac1{5!}\int_0^r(r-s)^5\Delta_p^{(5)}(s)\dd s,\\
 |R_{T,\dm}(p,r)|
 &\leq\frac{C(1+p^2)}{5!}\int_0^r(r-s)^5\dd s
 \leq C(1+p^2)r^6
 \qquad (0\leq r\leq\rho).
 \end{split}
\]
Taking $r=|z|$ proves \eqref{eq:R-bound}.
\end{proof}

\begin{lemma}[Large-momentum derivative bound]
\label{lem:rescaled-response}
For fixed $T>0$ and $\dm>0$, there exist $R,C>0$ and
$\rho\in(0,1/2)$ such that
\[
 |\Delta_p^{(5)}(r)|\leq C|\xi(p)|
 \qquad (p\in\R^d,\ |\xi(p)|\geq R,\ |r|\leq\rho).
\]
\end{lemma}

\begin{proof}
We write $\Delta_p$ as the inverse of a rescaled function and bound
the derivatives of that inverse uniformly in the scaling parameter.
For $x>0$, define $F_x\in C^\infty(\R;\R)$ by
\begin{equation}\label{eq:F-x-u}
 F_x(u):=
 \frac{u}{4\sqrt{1+u^2}}\left[
 \tanh\!\left(\frac{x\sqrt{1+u^2}+\dm}{2T}\right)
 +\tanh\!\left(\frac{x\sqrt{1+u^2}-\dm}{2T}\right)
 \right].
\end{equation}
For $x=|\xi(p)|>0$, this definition gives
\[
 F_x(u)=\frac{xu}{2}h_{T,\dm}(p,x^2u^2),
 \qquad
 \Delta_p(r)=xF_x^{-1}(r)\quad (|r|<1/2).
\]
Indeed, by the proof of Lemma~\ref{lem:fixed-gap-duality} and oddness,
the first expression is a strictly increasing bijection from $\R$ onto
$(-1/2,1/2)$, with positive derivative everywhere. Thus
$F_x^{-1}:(-1/2,1/2)\to\R$ is well-defined and smooth.
Substituting $u=\Delta_p(r)/x$ into the first identity and using
\eqref{eq:a-Delta} gives $F_x(\Delta_p(r)/x)=r$, which proves the
second identity. It remains to bound $(F_x^{-1})^{(5)}$ on an interval
independent of large $x$.

We first compare $F_x$ with $u/(2\sqrt{1+u^2})$. For
$x\geq\max\{1,\dm\}$,
\[
 \max_{0\leq k\leq5}\sup_{|u|\leq1}
 \left|\frac{\dd^k}{\dd u^k}
 \left(F_x(u)-\frac{u}{2\sqrt{1+u^2}}\right)\right|
 \leq C(1+x)^5e^{-(x-\dm)/T}\longrightarrow0
 \quad (x\to\infty).
\]
To see this, differentiating
$\tanh y-1=-2e^{-2y}/(1+e^{-2y})$ gives
\[
 \left|\frac{\dd^k}{\dd y^k}(\tanh y-1)\right|
 \leq Ce^{-2y}
 \qquad (y\geq0,\ k=0,\ldots,5).
\]
For $|u|\leq1$, the arguments of the hyperbolic tangents in
\eqref{eq:F-x-u} are at least $(x-\dm)/(2T)$.
The derivatives of $\sqrt{1+u^2}$ and $u/\sqrt{1+u^2}$ through
order five are bounded on this interval. The chain and product rules
therefore give the stated estimate, with at most five factors of $x$.

Since
\[
 \frac{\dd}{\dd u}\frac{u}{2\sqrt{1+u^2}}
 =\frac1{2(1+u^2)^{3/2}}>0,
\]
we may choose $R,C>0$ such that
\[
 F_x'(u)\geq C^{-1},
 \qquad |F_x^{(k)}(u)|\leq C
 \qquad (x\geq R,\ |u|\leq1,\ k=0,\ldots,5).
\]
Because $F_x(0)=0$, integration of the lower bound gives
$F_x(1)\geq C^{-1}$ and $F_x(-1)\leq-C^{-1}$.
Choose $0<\rho<\min\{1/2,C^{-1}\}$. Then
$F_x^{-1}([-\rho,\rho])\subset(-1,1)$ for every $x\geq R$,
so the preceding derivative bounds apply at $u=F_x^{-1}(r)$.

Differentiating $F_x(F_x^{-1}(r))=r$ first gives
\[
 (F_x^{-1})'(r)=\frac1{F_x'(F_x^{-1}(r))},
 \qquad |(F_x^{-1})'(r)|\leq C
 \quad (|r|\leq\rho).
\]
For $2\leq k\leq5$, the $k$th differentiated identity contains
the highest derivative of the inverse only in the term
\[
 F_x'(F_x^{-1}(r))(F_x^{-1})^{(k)}(r).
\]
Every other term is a product of derivatives of $F_x$ of order at
most $k$ and derivatives of $F_x^{-1}$ of order less than $k$.
Induction, using $F_x'\geq C^{-1}$, therefore bounds all inverse
derivatives through order five uniformly for $x\geq R$ and
$|r|\leq\rho$.
Finally, differentiating the identity $\Delta_p(r)=xF_x^{-1}(r)$
established above gives
\[
 |\Delta_p^{(5)}(r)|
 =x|(F_x^{-1})^{(5)}(r)|\leq Cx
 \qquad (x=|\xi(p)|\geq R,\ |r|\leq\rho),
\]
as claimed.
\end{proof}

For bounded continuous $f:\R^d\to\C$, define
\begin{equation}\label{eq:I-def}
 I_{T,\dm}(f)=\int_{\R^d}j_{T,\dm}(p)f(p)\dd p.
\end{equation}

\begin{lemma}[Low-temperature measure limit]\label{lem:T-zero-j}
Fix $\dm\in(0,\mub)$.  Let $f_T,f_0:\R^d\to\C$ be bounded continuous
functions such that $\|f_T-f_0\|_\infty\to0$ as $T\downarrow0$.  Then
\begin{equation}\label{eq:j-T-limit}
 \begin{split}
 \lim_{T\downarrow0}I_{T,\dm}(f_T)
 &=-\int_{\R^d}
 \frac{\1_{\{|p^2-\mub|>\dm\}}f_0(p)}{2|p^2-\mub|^3}\dd p\\
 &\quad+\frac1{4\dm^2}\left[
 (\mub-\dm)^{(d-2)/2}
 \int_{\Sph^{d-1}}f_0(\sqrt{\mub-\dm}\,\omega)\dd\omega\right.\\
 &\hspace{34mm}\left.+
 (\mub+\dm)^{(d-2)/2}
 \int_{\Sph^{d-1}}f_0(\sqrt{\mub+\dm}\,\omega)\dd\omega
 \right].
 \end{split}
\end{equation}
Moreover, $\sup_{0<T<T_1}\|j_{T,\dm}\|_{L^1}<\infty$ for sufficiently
small $T_1$, with a bound depending on $\dm$.
\end{lemma}

\begin{proof}
\emph{Roadmap.}  We preserve the cancellation between the two singular-looking
terms in $j$, split the radial variable into an interior region, a transition
layer, and a tail, and then use polar coordinates to obtain the limit in momentum space.

For $x\geq0$, put
\[
 D_T(x)=\tanh\!\left(\frac{x+\dm}{2T}\right)
 +\tanh\!\left(\frac{x-\dm}{2T}\right).
\]
For $x>0$, the two terms in \eqref{eq:j-def} must be kept together:
\begin{equation}\label{eq:j-combined}
 j_{T,\dm}(p)
 =\frac14\left(\frac{D_T'(x)}{x^2}-\frac{D_T(x)}{x^3}\right),
 \qquad x=|p^2-\mub|.
\end{equation}
This expression has a removable singularity at $x=0$.

We split the radial variable into three regions.  On $0\leq x\leq\dm/2$, note that
\[
 xD_T'(x)-D_T(x)=\int_0^x yD_T''(y)\dd y.
\]
Using that
$D_T$ is odd and $D_T''(0)=0$ we obtain
\[
 \left|\frac{D_T'(x)}{x^2}-\frac{D_T(x)}{x^3}\right|
 \leq\frac13\sup_{0\leq y\leq x}|D_T'''(y)|.
\]
For $y\leq\dm/2$, differentiation of the hyperbolic functions gives
\[
 |D_T'''(y)|\leq C T^{-3}e^{-\dm/(2T)}.
\]
Thus the contribution of this region tends to zero and is uniformly
integrable.  

On $\dm/2\leq x\leq3\dm/2$, one has
$D_T(x)\to2\1_{\{x>\dm\}}$ in $L^1$ and its derivative converges to a Dirac delta distribution
\[
 D_T'(x)=\frac1{2T}\left[
 \operatorname{sech}^2\!\left(\frac{x+\dm}{2T}\right)
 +\operatorname{sech}^2\!\left(\frac{x-\dm}{2T}\right)
 \right]
 \rightharpoonup2\delta(x-\dm)
\]
as finite measures.  
Hence, on this region, \eqref{eq:j-combined} converges as a signed measure to
\[
 -\frac{\1_{\{x>\dm\}}}{2x^3}\dd x
 +\frac1{2\dm^2}\delta(x-\dm).
\]
On $x\geq3\dm/2$, $D_T\to2$ and $D_T'\to0$ with exponential bounds, while
the limiting density $x^{-3}$ remains integrable after the radial change of
variables for $d\leq3$.  These estimates also give a uniform $L^1$ bound.

The weights in the radial integral are continuous near $x=\dm$, since
$0<\dm<\mub$.  In $d=1$, the weight also has an integrable singularity
at $x=\mub$, which lies away from $x=\dm$; the exponential bounds on
$D_T'$ and boundedness of $D_T$ control this region.  Thus the preceding
limits also hold after integration in momentum space.
Integration in polar coordinates gives the first term in
\eqref{eq:j-T-limit}.  For the boundary measure,
\[
 \int_{\BR^d}\delta(|p^2-\mub|-\dm)f_0(p)\dd p
 =\frac12\sum_{\sigma=\pm}
 (\mub+\sigma\dm)^{(d-2)/2}
 \int_{\Sph^{d-1}}f_0(\sqrt{\mub+\sigma\dm}\,\omega)\dd\omega.
\]
Multiplication by $1/(2\dm^2)$ gives the displayed coefficient.  Finally, the
uniform $L^1$ bound permits replacement of $f_T$ by $f_0$.
\end{proof}

\begin{lemma}[Small-imbalance asymptotic]\label{lem:delta-j}
Let $f_{\dm},f_*:\R^d\to\C$ be bounded continuous functions such that
$\|f_{\dm}-f_*\|_\infty\to0$ as $\dm\downarrow0$.  Denote the right side of \eqref{eq:j-T-limit} by
$I_{0,\dm}(f_{\dm})$.  Then
\begin{equation}\label{eq:j-delta-limit}
 \lim_{\dm\downarrow0}\dm^2 I_{0,\dm}(f_{\dm})
 =\frac{\mub^{(d-2)/2}}4
 \int_{\Sph^{d-1}}f_*(\sqrt{\mub}\,\omega)\dd\omega.
\end{equation}
\end{lemma}

\begin{proof}
The contribution of the first term in $I_{0,\dm}$ can be written as
\begin{equation}
- \dm^2\int_0^\infty \frac{\1_{\{|s-\mub|>\dm\}}}{4|s-\mub|^3}B_{\dm}(s)\dd s,
\end{equation}
where
\[
 B_{\dm}(s)=s^{(d-2)/2}
 \int_{\Sph^{d-1}}f_{\dm}(\sqrt{s}\,\omega)\dd\omega.
\]
On every compact neighborhood of $\mub$ contained in $(0,\infty)$, the functions $B_{\dm}$ are uniformly
bounded and converge uniformly to the analogous $B_*$.  Fix $0<\varepsilon<\mub/2$.  The part of the bulk
integral with $|s-\mub|\geq\varepsilon$ is $O(\dm^2)$, since
$|B_{\dm}(s)|\leq Cs^{(d-2)/2}$ and
$s^{(d-2)/2}|s-\mub|^{-3}$ is integrable on this set for $d\leq3$.
In the remaining part,
put $s=\mub+\dm v$.  Dominated convergence, with majorant $C|v|^{-3}$ on
$|v|>1$, gives
\[
 -\frac14\int_{|v|>1}\frac{B_*(\mub)}{|v|^3}\dd v
 =-\frac14B_*(\mub).
\]
The two boundary terms in $I_{0,\dm}$ contribute
\[
 \frac14\bigl(B_{\dm}(\mub-\dm)+B_{\dm}(\mub+\dm)\bigr)
 \longrightarrow\frac12B_*(\mub).
\]
Their sum is $B_*(\mub)/4$, which is \eqref{eq:j-delta-limit}.
\end{proof}

\begin{prop}[Negative quartic term at the linear threshold]
\label{prop:negative-quartic}
Under the hypotheses of Theorem~\ref{thm:main}, there exists
$\delta_{\mu,0}>0$ such that, for every $0<\dm<\delta_{\mu,0}$, there is
$T_0(\dm)>0$ with
\begin{equation}\label{eq:I-positive}
 I_{T,\dm}(|q_{T,\dm}|^4)>0
 \qquad(0<T<T_0(\dm)).
\end{equation}
At $\lambda=\llin(T,\dm)$,
\begin{equation}\label{eq:critical-trial-expansion}
 \mathcal E_{T,\dm,\llin}
 \left(\Gamma^{[t \alpha_{T,\dm}]}\right)
 =-2\llin^4
 I_{T,\dm}(|q_{T,\dm}|^4)t^4+O(t^6),
 \qquad t\in\R,\quad t\to0.
\end{equation}
In particular, this energy is negative for all sufficiently small nonzero $t$.
\end{prop}

\begin{proof}
\emph{Roadmap.}  For the first claim we combine the asymptotics of $q_{T,\dm}$ derived in Proposition~\ref{prop:BS} with the asymptotics of $I_{T,\dm}$ derived in the two preceding Lemmas as $T\to 0$ and $\dm \to 0$.  For the second claim we use Lemma~\ref{lem:quartic-expansion}, where we plug in the ground state of the Hessian operator and integrate over $p$.

By Proposition~\ref{prop:BS}, $q_{0,\dm}\to q_*$ uniformly.  Apply
Lemma~\ref{lem:delta-j} with $f_{\dm}=|q_{0,\dm}|^4$.  Equation
\eqref{eq:q-star-nonzero} gives
\[
 \dm^2I_{0,\dm}(|q_{0,\dm}|^4)
 \longrightarrow\frac{\mub^{(d-2)/2}}4
 \int_{\Sph^{d-1}}|q_*(\sqrt{\mub}\omega)|^4\dd\omega>0.
\]
Thus the zero-temperature expression is positive for small $\dm$.  For each
such $\dm$, Lemma~\ref{lem:T-zero-j} and the uniform convergence
$q_{T,\dm}\to q_{0,\dm}$ imply \eqref{eq:I-positive} for small $T$.

Let $\alpha=\alpha_{T,\dm}$ and $q=\Vop \alpha=q_{T,\dm}$.  The state
$\Gamma^{[t\alpha]}$ in \eqref{eq:optimizer-state} is measurable. 
Recall from Section~\ref{sec:overview} that 
\[
\mathcal{E}_{T,\dm,\llin}(\Gamma^{[t \alpha]})
=\mathcal{R}_{T,\dm,\llin}(t \alpha)
=\int_{\BR^d}G_{T,\dm}(p,t\alpha(p))\dd p
 -\llin t^2\langle\alpha,\Vop\alpha\rangle.
\]
Since $\alpha\in L^\infty$, \eqref{eq:G-expansion} applies whenever
$|t|\|\alpha\|_\infty\leq\rho$. 
The error term is integrable and thus of order $O(t^6)$: $\alpha\in L^\infty$ and
\eqref{eq:K-growth} imply
\[
 \int(1+p^2)|\alpha|^6
 \leq\|\alpha\|_\infty^4\int(1+p^2)|\alpha|^2<\infty.
\]

The quadratic term cancels by \eqref{eq:linear-equation}, and
$K_{T,\dm}\alpha=\llin q$ gives
\[
 \int K_{T,\dm}^4j_{T,\dm}|\alpha|^4
 =\llin^4\int j_{T,\dm}|q_{T,\dm}|^4 = \llin^4 I_{T,\dm}(|q_{T,\dm}|^4).
\]
This proves
\eqref{eq:critical-trial-expansion}.
\end{proof}

\subsection{Proof of the main theorem}\label{eq:conclusion}
Now we have all ingredients to conclude the main result.
\begin{proof}[Proof of Theorem~\ref{thm:main}]
Under the assumptions on $V$ in \eqref{eq:V-assumptions}, the operator $L_{T,\dm}$ is compact and self-adjoint for all $T,\dm>0$ by Lemma~\ref{lem:compact-sandwich}.
By Propositions~\ref{prop:BS} and \ref{prop:negative-quartic}, there is
$\delta_{\mu,0}\in(0,\mub)$ such that, for every $0<\dm<\delta_{\mu,0}$,
there is $T_0(\dm)>0$ for which $b_{T,\dm}>0$ and the critical trial states
have negative energy whenever $0<T<T_0(\dm)$ and $t\neq0$ is sufficiently small.
As discussed in Remark \ref{rmk:topology}, the trial states
converge to $\Gamma_0$ in the topology specified in Section~\ref{sec:model-main}.
Thus the normal state is not a local minimizer at $\lambda=\llin(T,\dm)$.
For $\lambda>\llin$, the same critical mode has quadratic coefficient
\[
 \langle\alpha_{T,\dm},(K_{T,\dm}-\lambda\Vop)\alpha_{T,\dm}\rangle
 =(\llin-\lambda)\langle\alpha_{T,\dm},\Vop\alpha_{T,\dm}\rangle<0,
\]
because $\llin\langle\alpha_{T,\dm},\Vop\alpha_{T,\dm}\rangle
=\langle\alpha_{T,\dm},K_{T,\dm}\alpha_{T,\dm}\rangle>0$.
This also proves local instability above $\llin$; compare
\cite[Theorem~1]{FreijiHainzlSeiringer2012}.
After rescaling the critical mode, we obtain
$\alpha_0\in\Da$, $|\alpha_0|\leq1/2$, with
$\mathcal R_{T,\dm,\llin}(\alpha_0)<0$.
Proposition~\ref{prop:variational-first-order} now gives the remaining conclusions.
\end{proof}

\section*{AI declaration}
The authors used GPT-5.5 Pro (2026) to explore proof strategies, carry out computations and prepare a draft of the manuscript. The authors verified and substantially revised the generated content. The authors are responsible for the entire content of the manuscript.

\section*{Acknowledgments}
The authors thank Andreas Deuchert and Robert Seiringer for helpful discussions.
This research was supported by the DFG under grant TRR 352 – Project-ID 470903074 (M.L.\ and B.R.) and FOR 5413 – Project-ID 465199066 (M.L.), as well as by the European Union through ERC Starting Grant MathQuantProp, Grant Agreement 101163620 (M.L.) and the ERC Starting Grant MaTCh, grant agreement n.~101117299 (B.R.).\footnote{Views and opinions expressed are however those of the authors only and do not necessarily
	reflect those of the European Union or the European Research Council Executive Agency. Neither
	the European Union nor the granting authority can be held responsible for them.} B.R.\ also warmly acknowledges the GNFM (Gruppo Nazionale per la Fisica Matematica) - INDAM.

\appendix

\section{Quasi-free origin of the reduced model}\label{sec:model_justification}

This appendix explains two models whose translation-invariant BCS functional reduces to \eqref{eq:BCSfunctional}, with an explicit coupling normalization in the singlet case.
The general formalism of quasi-free states and generalized one-particle density matrices is described in \cite{BachFrohlichJonsson2009,BachLiebSolovej1994}.
For almost every $p\in\R^d$, let
\[
 \gamma(p)=\gamma(p)^*\in\C^{2\times2},
 \qquad \alpha(p)\in\C^{2\times2},
 \qquad \alpha(p)^T=-\alpha(-p).
\]
The generalized one-particle density matrix is
\begin{equation}\label{eq:general-Gamma-appendix}
 \boldsymbol\Gamma(p)=
 \begin{pmatrix}
 \gamma(p)&\alpha(p)\\
 \alpha(p)^*&1-\overline{\gamma(-p)}
 \end{pmatrix},
 \qquad \boldsymbol\Gamma(p)\in\C^{4\times4},\quad 0\leq\boldsymbol\Gamma(p)\leq1.
\end{equation}

Consider a BCS interaction which acts only in the two ordered
opposite-spin channels.  In spin space take the interaction matrix
\begin{equation}\label{eq:opposite-spin-projector}
 Q_{\mathrm{op}}
 =\frac12\left(
 |\uparrow\downarrow\rangle\langle\uparrow\downarrow|
 +|\downarrow\uparrow\rangle\langle\downarrow\uparrow|
 \right)\in\C^{4\times4}.
\end{equation}
With the sign convention of the main text, the pairing contribution per unit
volume is
\begin{equation}\label{eq:appendix-pair-energy}
 -\frac\lambda2\int_{\R^d}V(x)
 \left(|\alpha_{\uparrow\downarrow}(x)|^2
 +|\alpha_{\downarrow\uparrow}(x)|^2\right)\dd x.
\end{equation}
Here the position-space components are the inverse Fourier transforms of
the momentum-space components.  Since $V$ is even and
$\alpha_{\downarrow\uparrow}(x)=-\alpha_{\uparrow\downarrow}(-x)$, the two
integrals in \eqref{eq:appendix-pair-energy} are equal.  Hence the interaction is
$-\lambda\int V|\alpha_{\uparrow\downarrow}|^2$, exactly as in
\eqref{eq:functional}.  

We next justify the absence of equal-spin pairings.  
The replacement
\[
 \boldsymbol\Gamma(p)\longmapsto
 \frac12\bigl(\boldsymbol\Gamma(p)+U\boldsymbol\Gamma(p)U^*\bigr),
 \qquad U=\operatorname{diag}(1,-1,-1,1)\in\C^{4\times4},
\]
sets the off-diagonal entries of $\gamma$ and the diagonal entries of
$\alpha$ to zero.  It preserves the constraint, the kinetic energy, and the
interaction, and does not increase the negative entropy term by convexity
\cite[Lemma 3.1]{BachLiebSolovej1994}.
After this replacement, write
$\alpha(p)=\alpha_{\uparrow\downarrow}(p)$ and use
$\alpha_{\downarrow\uparrow}(p)=-\alpha(-p)$.  The matrix has the form
\begin{equation}\label{eq:reduced-state}
 \begin{pmatrix}
 \gamma_+(p)&0&0&\alpha(p)\\
 0&\gamma_-(p)&-\alpha(-p)&0\\
 0&-\overline{\alpha(-p)}&1-\gamma_+(-p)&0\\
 \overline{\alpha(p)}&0&0&1-\gamma_-(-p)
 \end{pmatrix}.
\end{equation}
Compose the second and third components with the reflection $p\mapsto-p$.  This is a unitary transformation.  
Since the dispersion $p^2$ is even, it preserves the kinetic energy.
It also preserves the entropy and the pairing quadratic form.  
After relabeling $\gamma_-(-p)$ as $\gamma_-(p)$, one obtains exactly
\eqref{eq:block-state}.  

Alternatively, consider an interaction supported on the spin-singlet
subspace.  Its spin component is the $SU(2)$-invariant projection
\begin{equation}
  Q_{\rm s}=\frac12(  |\uparrow\downarrow\rangle-|\downarrow\uparrow\rangle)(\langle\uparrow\downarrow|-\langle\downarrow\uparrow|)\in\C^{4\times4}.
\end{equation}
The interaction energy then is
\begin{equation}\label{eq:appendix-pair-energy-2}
 -\frac\lambda2\int_{\R^d}V(x)
|\alpha_{\uparrow\downarrow}(x)
 -\alpha_{\downarrow\uparrow}(x)|^2\dd x
 =-\frac\lambda2\int_{\R^d}V(x)
|\alpha_{\uparrow\downarrow}(x)
 +\alpha_{\uparrow\downarrow}(-x)|^2\dd x.
\end{equation}
The same matrix averaging restricts the state to the form
\eqref{eq:reduced-state}.  Replacing this state by
$\frac12\boldsymbol\Gamma(p)+\frac12\boldsymbol\Gamma(-p)$ preserves the
kinetic and singlet interaction energies and does not increase the negative
entropy term.  The resulting state has the form \eqref{eq:block-state}, with
$\gamma_\pm(p)=\gamma_\pm(-p)$ and $\alpha(p)=\alpha(-p)$.
For such states, \eqref{eq:appendix-pair-energy-2} equals
$-2\lambda\int_{\R^d}V(x)|\alpha(x)|^2\dd x$.
Thus the reduced functional is \eqref{eq:BCSfunctional} with coupling
$2\lambda$; equivalently, replacing $\lambda$ by $\lambda/2$ in
\eqref{eq:appendix-pair-energy-2} gives precisely the normalization of the
main text.  With this normalization, the main theorem also applies to the
even subspace if the eigenfunction corresponding to the largest eigenvalue
of $\Vop_{\mub}$ is even.

\section{Comparison with Freiji--Hainzl--Seiringer}\label{app:temperature}

\subsection{The set of first-order transition temperatures}
Freiji--Hainzl--Seiringer obtain spectral bounds on the transition
and describe an additional normal--superfluid boundary, between their
inner curve $T_{\dm}^{\mathrm i}$ and improved weak-coupling bound
$T_{\dm}^{\mathrm g}$, whose existence
is suggested by numerical calculations but is not established by their
analysis \cite[Section~2.1.1]{FreijiHainzlSeiringer2012}.
This is the first-order line~III of Caldas--Mota
\cite[Fig.~2]{caldas_temperature_2008}, (which is different from region~III in
\cite[Fig.~1]{FreijiHainzlSeiringer2012}).
Theorem~\ref{thm:main} establishes the corresponding first-order
coexistence for the potentials and the low-temperature,
small-imbalance regime considered here.

Assume the hypotheses of Theorem~\ref{thm:main}.  For fixed
$\lambda>0$ and $\dm\in(0,\delta_{\mu,0})$, define the temperature sets
\begin{equation}\label{eq:B-short-temperature-sets}
\begin{aligned}
 \mathcal T_{\dm}^{\mathrm i}(\lambda)
 &:=\bigl\{T\in(0,T_0(\dm)):
       \inf\sigma(K_{T,\dm}-\lambda\Vop)=0\bigr\},\\
 \mathcal T_{\dm}^{\mathrm{fo}}(\lambda)
 &:=\bigl\{T\in(0,T_0(\dm)):
       \lfo(T,\dm)=\lambda\bigr\}.
\end{aligned}
\end{equation}
The spectra are taken on $L^2(\R^d)$. $\mathcal T_{\dm}^{\mathrm i}(\lambda)$ is defined in this way in \cite{FreijiHainzlSeiringer2012} and the condition $\inf\sigma(K_{T,\dm}-\lambda\Vop)=0$ can be reformulated using our \eqref{eq:lambda-lin}
as $\llin(T,\dm)=\lambda$.
In $d=3$, it describes the restriction of the inner temperature
curve in \cite[Eq.~(2.7)]{FreijiHainzlSeiringer2012}, with their
potential replaced by $-V$ to match our sign convention.

The second set, $ \mathcal T_{\dm}^{\mathrm{fo}}(\lambda)$, is introduced here. For $0<T<T_0(\dm)$, Theorem~\ref{thm:main} establishes that it is exactly the set of coexistence
temperatures: 
\begin{equation}\label{eq:B-short-coexistence}
 T\in\mathcal T_{\dm}^{\mathrm{fo}}(\lambda)
 \quad\Longleftrightarrow\quad
 \exists\,\Gamma_*\in\mathcal D:\quad
 (\Gamma_*)_{14}\ne0,\qquad
 \mathcal E_{T,\dm,\lambda}(\Gamma_*)
 =\inf_{\Gamma\in\mathcal D}\mathcal E_{T,\dm,\lambda}(\Gamma)=0,
\end{equation}
recalling that the normal state also has energy zero. 
Moreover, in the regime of validity of Theorem~\ref{thm:main}, we have the strict inequality
$\lfo(T,\dm)<\llin(T,\dm)$.  
Thus, for every  $T\in\mathcal T_{\dm}^{\mathrm{fo}}$, we have 
$\inf\sigma(K_{T,\dm}-\lambda\Vop)>0$: the paired and normal
states are global minimizers while the normal state remains a strict
local minimizer.  For the in-between region of temperatures
\begin{equation}\label{eq:B-short-metastability}
 T\in\bigl\{s\in(0,T_0(\dm)):
       \lfo(s,\dm)<\lambda<\llin(s,\dm)\bigr\}
\end{equation}
every global minimizer is paired while
the normal state remains a strict local minimizer.

The nonzero pairing at coexistence and the kink of
$e_{T,\dm}(\lambda)$ as the coupling varies, proved in
\eqref{eq:one-sided-slopes-main}, establish a first-order transition,
rather than merely the existence of a nonzero gap-equation solution.

Consequently, $\mathcal T_{\dm}^{\mathrm{fo}}(\lambda)$ supplies the
coexistence criterion missing from the spectral bounds of
\cite{FreijiHainzlSeiringer2012}, in the regime of Theorem~\ref{thm:main}.
This gives a rigorous counterpart, with coupling as the transition parameter, of the line~III numerically found in \cite{caldas_temperature_2008}, for the present potentials.
The definition of transition temperatures as sets in \eqref{eq:B-short-temperature-sets} is conveniently general. Locally realizing these sets as curves in the $(\delta_\mu,T)$ phase diagram requires sufficient regularity of $\lfo$ and a nonvanishing derivative with respect to $T$ to apply the implicit function theorem. We expect these assumptions to hold generically, but we do not aim to further study the set $\mathcal T_{\dm}^{\mathrm{fo}}$ analytically here.

\subsection{Sign change of the quartic in a low-temperature scaling limit}\label{sec:simultaneous_scaling}

The main theorem uses the sequential limit $T\downarrow0$ followed by $\dm\downarrow0$, leading to a negative quartic coefficient.
If the order of the limits is reversed, one obtains the usual spin-balanced case, where the quartic coefficient is positive.
By considering the simultaneous scaling $\dm=\kappa T$ we can pinpoint where the quartic sign changes.
This simultaneous scaling has been studied in \cite{FreijiHainzlSeiringer2012}.
For the phase diagram in Figure~\ref{fig:phases} that work proves
superfluidity in region I and normality in regions III and IV, while phase II is unknown; its conclusion in
region III requires additional hypotheses on the potential including radiality and a nonnegative Fourier transform \cite[Theorem~4]{FreijiHainzlSeiringer2012}.  
The numerical discussion following that theorem shows that the tip of region II is at $\dm/T\simeq1.91$.

\begin{figure}
    \centering
\begin{tikzpicture}[x=5.6cm,y=6.5cm,
    boundary/.style={black,line width=1.15pt,line cap=round,line join=round}]

  \fill[teal!40]
    (1.085108,.51)
    .. controls (1.085990,.500310) and (1.086339,.490155) .. (1.086,.48)
    .. controls (1.092,.335) and (.94,.205) .. (.895,.105)
    .. controls (.877,.065) and (.877,.02) .. (.877,0)
    -- (.922,0)
    .. controls (.922,.045) and (.938,.09) .. (.955,.125)
    .. controls (.99,.225) and (1.132,.35) .. (1.102,.455)
    .. controls (1.099,.47) and (1.090,.50) .. (1.085108,.51)
    -- cycle;
    
  \draw[black,line width=.45pt] (0,0) -- (1.97,0);
  \draw[black,line width=.45pt] (0,0) -- (0,1.02);
  \foreach \x in {0.1,0.2,...,1.9} {
    \draw[line width=.35pt] (\x,0) -- ++(0,.011);
  }
  \foreach \x in {0.5,1,1.5} {
    \draw[line width=.5pt] (\x,0) -- ++(0,.02);
    \node[anchor=north] at (\x,-.013) {$\x$};
  }
  \foreach \y in {0.05,0.10,...,0.95} {
    \draw[line width=.35pt] (0,\y) -- ++(.012,0);
  }
  \foreach \y in {0.2,0.4,0.6,0.8,1.0} {
    \draw[line width=.5pt] (0,\y) -- ++(.023,0);
    \node[anchor=east] at (-.015,\y) {$\y$};
  }
  \node[anchor=south] at (0,1.04) {$T$};
  \node[anchor=west] at (2.015,0) {$\delta_\mu$};

  \draw[boundary] (0,1)
    .. controls (.34,1) and (.79,.92) .. (.96,.728)
    .. controls (1.16,.48) and (1.53,.23) .. (1.94,.112);

  \draw[boundary] (.96,.728)
    .. controls (1.025,.65) and (1.058,.61) .. (1.075,.558)
    .. controls (1.12,.34) and (1.43,.065) .. (1.735,0);

  \draw[boundary] (.96,.728)
    .. controls (1.03,.66) and (1.085,.57) .. (1.086,.48)
    .. controls (1.092,.335) and (.94,.205) .. (.895,.105)
    .. controls (.877,.065) and (.877,.02) .. (.877,0);

  \fill (.96,.728) circle[radius=1.35pt];
  \fill (1.075,.558) circle[radius=1.35pt];
  \node[anchor=west] at (.975,.814)
    {$\delta_\mu/T=\cosh^{-1}(2)$};
  \node[anchor=west] at (1.14,.605)
    {$\delta_\mu/T=1.91$};

  \node at (.50,.50) {I};
  \node at (1.15,.18) {II};
  \node at (1.45,.245) {III};
  \node at (1.70,.395) {IV};
\end{tikzpicture}    \caption{Schematic phase regions adapted from \cite[Figure~1]{FreijiHainzlSeiringer2012}.  
Both axes are measured in units of the spin-balanced critical temperature. 
Region I is a superconducting phase, III and IV are normal phases, while phase II is unknown \cite{FreijiHainzlSeiringer2012}.
The shaded strip schematically illustrates the local extension of superconductivity beyond the linear-instability boundary discussed in Appendix B.2.}
    \label{fig:phases}
\end{figure}

Recall that at critical coupling $\llin$, the leading order of the energy expansion \eqref{eq:critical-trial-expansion} around the normal state is given by the quartic coefficient proportional to $-I_{T,\kappa T}(|q_{T,\kappa T}|^4)$.
Similarly to Proposition~\ref{prop:BS}, also in this scaling $q_{T,\kappa T}\to q_{*}$ uniformly as $T\downarrow 0$. (This follows considering the Fermi-shell decomposition for $A_{T,\kappa T}$, which for $d=3$ is discussed in \cite[Thm 2]{FreijiHainzlSeiringer2012} and can be easily adapted from \cite[Lemma 3.4]{HenheikLauritsenRoos2024} in the case $d=1,2$.)
The following Proposition gives the limit of the quartic term if we set $f_T=|q_{T,\kappa T}|^4, f_0=|q_*|^4$.
\begin{prop}[Simultaneous low-temperature scaling]
\label{prop:simultaneous-scaling}
Let $d\in\{1,2,3\}$, $\mub>0$, and $\kappa\geq0$ be fixed.
Let $f_T,f_0:\R^d\to\C$ be bounded continuous
functions such that $\|f_T-f_0\|_\infty\to0$ as $T\downarrow0$. 
For $s>0$, set
\begin{equation}\label{eq:B-f-mu}
 B_f(s)=s^{(d-2)/2}
 \int_{\Sph^{d-1}}f(\sqrt{s}\,\omega)\dd\omega.
\end{equation}
Then
\begin{equation}\label{eq:simultaneous-polygamma-limit}
 \lim_{T\downarrow0}T^2
 I_{T,\kappa T}(f_T)
 =\frac{B_{f_0}(\mub)}{16\pi^2}
 \operatorname{Re}\psi^{(2)}
 \left(\frac12+\frac{i\kappa}{2\pi}\right),
\end{equation}
where $\psi^{(2)}(z)=\frac{\dd^3}{\dd z^3}\log\Gamma(z)$ for
$\operatorname{Re}z>0$ is the polygamma function of order $2$ and $\Gamma$
is the Euler gamma function.  Moreover, 
\begin{equation}\label{eq:polygamma-endpoint-asymptotics}
 \operatorname{Re}\psi^{(2)}\!\left(\frac12\right)
 =-14\zeta(3)<0,
 \qquad
 \operatorname{Re}\psi^{(2)}
 \left(\frac12+\frac{i\kappa}{2\pi}\right)
 =\frac{4\pi^2}{\kappa^2}+O(\kappa^{-4})
\end{equation}
as $\kappa\to\infty$. 
\end{prop}

Interestingly, this agrees with the polygamma representation of the fourth-order coefficient in the constant-density-of-states contact model \cite[Eq.~(2.92)]{CasalbuoniNardulli2004}.
Numerical evaluation indicates a unique positive zero
\begin{equation}\label{eq:kappa-star}
 \operatorname{Re}\psi^{(2)}
 \left(\frac12+\frac{i\kappa_*}{2\pi}\right)=0,
 \qquad
 \kappa_*\simeq1.910668.
\end{equation}
This agrees with the numerically identified tip of region II in
\cite{FreijiHainzlSeiringer2012}.
Numerical evaluation indicates that the limiting quartic coefficient is positive for $\kappa<\kappa_*$ and negative for $\kappa>\kappa_*$.
For each fixed $\kappa$ for which this limiting coefficient is negative, and sufficiently small $T$, the negative quartic coefficient at the linear-instability boundary produces a paired trial state with strictly lower energy than $\Gamma_0$; continuity then extends this conclusion to nearby parameters on the side where the normal state is locally stable.
Thus, the superconducting phase extends locally into region II, as illustrated schematically by the shaded strip in Figure~\ref{fig:phases}.

\begin{proof}[Proof of Proposition~\ref{prop:simultaneous-scaling}.]
The proof has three steps.  We first discard the contribution away from
the Fermi surface and rescale the remaining integral.  We then evaluate the
limiting integral by Fourier transformation.  Finally, standard formulas
for the polygamma function give the endpoint value and asymptotics.

\emph{Step 1: rescaling near the Fermi surface.}

Split the momentum integral into $|p^2-\mub|<\mub/2$ and its complement.  For $x=|p^2-\mub|\geq\mub/2$, formula \eqref{eq:j-combined}, with
$\dm=\kappa T$, gives for all sufficiently small $T$
\begin{equation}\label{eq:simultaneous-off-shell-bound}
 |j_{T,\kappa T}(p)|
 \leq\frac{C}{x^3}
 +\frac{C}{Tx^2}e^{-x/(2T)}.
\end{equation}
For $x>0$,
\[
 \frac1T e^{-x/(2T)}\leq\frac{2}{e\,x},
\]
as follows by maximizing $y e^{-y}$ with $y=x/(2T)$.  Hence the right side
of \eqref{eq:simultaneous-off-shell-bound} is bounded by $C x^{-3}$,
uniformly in $T$.  This is integrable in $p$ on the complement of the shell:
there is no finite-momentum singularity because $x\geq\mub/2$, and it is
$O(|p|^{-6})$ at infinity.  Since $T^2j_{T,\kappa T}(p)\to0$ pointwise there,
dominated convergence shows that
\begin{equation}\label{eq:simultaneous-off-shell-vanishing}
 T^2\int_{|p^2-\mub|\geq\mub/2}
 j_{T,\kappa T}(p)f_T(p)\dd p\longrightarrow0.
\end{equation}

On every compact subinterval of $(0,\infty)$ 
the function $B_f$ is bounded and continuous 
and $B_{f_T}\to B_{f_0}$ uniformly as $T\to 0$.
Integrating over the angular variables and substituting $s=\mub+Tx$ gives the exact identity
\begin{equation}
 T^2\int_{|p^2-\mub|<\mub/2}
       j_{T,\kappa T}(p)f(p)\dd p
       =\frac12\int_{-\mub/2T}^{\mub/2T}
 \mathfrak j_\kappa(x) B_f(\mub+Tx)\dd x,
 \label{eq:simultaneous-shell-rescaling}
\end{equation}
where
\begin{equation}\label{eq:j-kappa-definition}
 \begin{aligned}
 \mathfrak j_\kappa(x)
 &=\frac14\left(
 \frac{D_\kappa'(|x|)}{|x|^2}
 -\frac{D_\kappa(|x|)}{|x|^3}\right),\qquad x\neq0,\\
 D_\kappa(x)&=
 \tanh\!\left(\frac{x+\kappa}{2}\right)
 +\tanh\!\left(\frac{x-\kappa}{2}\right).
 \end{aligned}
\end{equation}
Here $D_\kappa\in C^\infty(\R;\R)$ is odd.  Since
$D_\kappa(0)=D_\kappa''(0)=0$, Taylor's theorem gives
\[
 xD_\kappa'(x)-D_\kappa(x)
 =\frac13D_\kappa'''(0)x^3+O(x^5)
 \qquad(x\to0).
\]
Thus \eqref{eq:j-kappa-definition} extends continuously to $x=0$ with
$\mathfrak j_\kappa(0)=D_\kappa'''(0)/12$.  For
$x\geq1$, $D_\kappa(x)$ is bounded and $D_\kappa'(x)$ decays exponentially,
so
\begin{equation}\label{eq:j-kappa-majorant}
 |\mathfrak j_\kappa(x)|\leq
 \frac{C}{1+|x|^3},
 \qquad x\in\R.
\end{equation}
The integrable majorant \eqref{eq:j-kappa-majorant}, the convergence $B_{f_T}(\mub +Tx)\to B_{f_0}(\mub)$, and \eqref{eq:simultaneous-shell-rescaling} now yield
\begin{equation}\label{eq:J-kappa-integral}
 \lim_{T\downarrow0}T^2I_{T,\kappa T}(f_T)
 =\frac{B_{f_0}(\mub)}2
 \int_{\R}\mathfrak j_\kappa(x)\dd x.
\end{equation}
\emph{Step 2: evaluation of the limiting integral.}
Integration by parts on $(0,\infty)$ gives
\begin{equation}
 \int_{\R}\mathfrak j_\kappa(x)\dd x
 =\frac12\int_0^\infty
 \frac{xD_\kappa'(x)-D_\kappa(x)}{x^3}\dd x
 =\frac14\int_0^\infty\frac{D_\kappa''(x)}x\dd x.
 \label{eq:j-kappa-integration-by-parts}
\end{equation}
The boundary terms vanish because $xD_\kappa'(x)-D_\kappa(x)$ is
$O(x^3)$ at zero and bounded at infinity.  
For $r>0$ and $x>0$, we have 
\begin{equation}\label{eq:Abel-one-over-x}
 \int_0^\infty e^{-rs}\sin(sx)\dd s
 =\frac{x}{r^2+x^2}\xrightarrow[r\downarrow0]{}\frac1x.
\end{equation}
Because $D_\kappa''(x)=O(x)$ at zero and decays exponentially at infinity,
$|D_\kappa''(x)|/x$ is integrable on $(0,\infty)$.  Thus dominated
convergence and Fubini's theorem give
\begin{equation}
 \frac14\int_0^\infty\frac{D_\kappa''(x)}x\dd x
 =\lim_{r\downarrow0}\frac14\int_0^\infty e^{-rs}
       \left(\int_0^\infty D_\kappa''(x)\sin(sx)\dd x\right)\dd s
=-\lim_{r\downarrow0}\frac{\sqrt{2\pi}}{8}\int_0^\infty e^{-rs}
       s\widehat{D_\kappa'}(s)\dd s,
 \label{eq:Abel-Fourier-step}
\end{equation}
where the Fourier transform is unitary, as in the main text.  In the second
step we used that $D_\kappa''$ is odd and $\widehat{D_\kappa''}(s)=is\widehat{D_\kappa'}(s)$.
The Fourier transform formula \cite[Eq.~3.982.1]{GradshteynRyzhik2014} gives
\begin{equation}\label{eq:sech-square-Fourier}
 \int_\R e^{-isx}\sech^2(x/2)\dd x
 =\frac{4\pi s}{\sinh(\pi s)} \quad \text{and thus}\quad  \widehat{D_\kappa'}(s)
 =\frac{4\pi s\cos(\kappa s)}{\sqrt{2\pi} \sinh(\pi s)}.
\end{equation}
Combining this with \eqref{eq:Abel-Fourier-step} and \eqref{eq:j-kappa-integration-by-parts} gives
\begin{equation}\label{eq:j-kappa-Fourier-integral}
 \int_{\R}\mathfrak j_\kappa(x)\dd x
 =-\frac\pi2\int_0^\infty
 \frac{s^2\cos(\kappa s)}{\sinh(\pi s)}\dd s.
\end{equation}
For $\operatorname{Re}z>0$, the polygamma integral representation is
\cite[Section~6.4]{abramowitz1965}
\begin{equation}\label{eq:polygamma-integral-representation}
 \psi^{(2)}(z)
 =-\int_0^\infty\frac{t^2e^{-zt}}{1-e^{-t}}\dd t.
\end{equation}
Take $z=1/2+i\kappa/(2\pi)$, take real parts, and put $t=2\pi s$.
Equations \eqref{eq:j-kappa-Fourier-integral} and
\eqref{eq:polygamma-integral-representation} give
\[
 \int_{\R}\mathfrak j_\kappa(x)\dd x
 =\frac1{8\pi^2}\operatorname{Re}\psi^{(2)}
 \left(\frac12+\frac{i\kappa}{2\pi}\right).
\]
Together with \eqref{eq:J-kappa-integral}, this proves
\eqref{eq:simultaneous-polygamma-limit}.

\emph{Step 3: endpoint value and asymptotics.}
The formula \cite[Eq.~6.4.4]{abramowitz1965} gives
$\psi^{(2)}\left(\frac12\right)=-14\zeta(3)<0$, where $\zeta$ is the Riemann zeta function.   
Finally, the expansion \cite[Eq.~6.4.13]{abramowitz1965}
\[
 \psi^{(2)}(z)
 =-z^{-2}-z^{-3}-\frac12z^{-4}+O(|z|^{-6})
 \qquad(|z|\to\infty,\ |\arg z|\leq\pi-\delta)
\]
valid for every fixed $\delta>0$, with $z=1/2+i\kappa/(2\pi)$, gives
the large-$\kappa$ asymptotics and completes the proof.
\end{proof}

\end{document}